\RequirePackage{rotating}
\documentclass[fleqn,usenatbib]{mnras}

\usepackage{newtxtext,newtxmath}
\usepackage[T1]{fontenc}
\usepackage{ae,aecompl}

\usepackage{graphicx}	% Including figure files
\usepackage{amsmath}	% Advanced maths commands
\usepackage{amssymb}	% Extra maths symbols
\usepackage{rotating}
\usepackage{pdflscape}	% Landscape pages
\usepackage{hyperref}

\def\arcsec{^{\prime\prime}}

\newcommand\Fig[1]{Fig.~\ref{#1}}
\newcommand\Tab[1]{Table~\ref{#1}}
\newcommand\Sec[1]{Sec.~\ref{#1}}
\title[VIMOS Bulge variables]{New variable stars in the Galactic Bulge. I. The bright regime\thanks{Based on observations collected at the European Organisation for Astronomical Research in the Southern Hemisphere under ESO programmes 091.D-0489(A) and 093.D-0522(A).}}
\author[N. Kains et al.]{N. Kains$^{1}$,
A. Calamida$^{1}$,
M. Rejkuba$^{2}$,
A. Bhardwaj$^{2,3}$,
L. Inno$^{4}$,
K. C. Sahu$^{1}$,
\newauthor
M. Zoccali$^{5,6}$,
G. Bono$^{7,8}$,
F. Surot$^{2}$,
J. Anderson$^{1}$,
and S. Casertano$^{1}$
\\
$^{1}$Space Telescope Science Institute, 3700 San Martin Drive, Baltimore, MD 21218, USA\\
$^{2}$European Southern Observatory, Karl-Schwarzschild Stra{\ss}e 2, 85748 Garching bei M\"{u}nchen, Germany\\
$^{3}$Department of Physics \& Astrophysics, University of Delhi, Delhi, 110007, India\\
$^{4}$Max-Planck Institute for Astronomy, K\"{o}nigstuhl 17, 69117 Heidelberg, Germany\\
$^{5}$Pontificia Universidad Cat\'{o}lica de Chile, Instituto de Astrofisica, Av. Vicu\~{n}a Mackenna 4860, Santiago, Chile\\
$^{6}$Millennium Institute of Astrophysics, Av. Vicu\~na Mackenna 4860, 782-0436 Macul, Santiago, Chile\\
$^{7}$Universit\`{a} di Roma Tor Vergata, Via della Ricerca Scientifica 1, 00133 Roma, Italy\\
$^{8}$INAF-Osservatorio Astronomico di Roma, Via Frascati 33, 00040 Monteporzio Catone, Italy
}

\date{}

\pubyear{2018}

\begin{document}
\label{firstpage}
\pagerange{\pageref{firstpage}--\pageref{lastpage}}
\maketitle

\begin{abstract}
We report the detection of 1143 variable stars towards the Galactic bulge, including 320 previously uncatalogued variables, using time-series photometry extracted from data obtained with the VIMOS imager at the Very Large Telescope. Observations of the Sagittarius Window Eclipsing Extrasolar Planet Search (SWEEPS) field in the Galactic Bulge were taken over 2 years between March and October at a cadence of $\sim$ 4 days, enabling the detection of variables with periods up to $\sim$100 days. Many of these were already known, but we detected a significant number of new variables, including 26 Cepheids, a further 18 Cepheid candidates, and many contact binaries. Here we publish the catalog of the new variables, containing  coordinates, mean magnitudes as well as periods and classification; full light curves for these variables are also made available electronically.
\end{abstract}


%%%%%%%%%%%%%%%%%%%%%%%%%%%%%%%%%%%%%%%%%%%%%%%%%%

%%%%%%%%%%%%%%%%% BODY OF PAPER %%%%%%%%%%%%%%%%%%

\section{Introduction}

Variable stars have played a crucial role in astronomy for decades, enabling us to study not only the properties and structure of stars themselves, but also those of their environments, the stellar clusters or galaxies they inhabit. Variability surveys account for some of the earliest large-scale studies in modern astronomy, exemplified for instance by the work of Alexander Roberts \citep[e.g.][]{roberts1895}, Henrietta Swan Leavitt \citep[e.g.][]{leavitt04}, Annie Jump Cannon \citep[e.g.][]{cannon00}, Harlow Shapley \citep[e.g.][]{shapley19}, and Edward Pickering \citep[e.g.][]{pickering16}, to name only a few. These studies unearthed a wide variety of variable stars, spanning a huge range of brightness, color, period, light curve morphology. Period-luminosity relations \citep{leavitt08, leavitt12}, and, more recently, period-luminosity-metallicity relations for some types of variables \citep[e.g.][]{marconi15}, have enabled us to gain deep insights into the structure of the Milky Way \citep[e.g.\, ][]{payne-gaposchkin1952,sesar2013,dekany2013,dekany2015, pietrukowicz2015, catchpole2016,bhardwaj2017} and other nearby galaxies \citep[e.g.\ ][]{moretti2014, ripepi2017}, but also into the evolution of the Universe, via constraints on the Hubble Constant $H_0$ \citep[e.g.\, ][]{freedman01,riess16}.

Binary systems are also a rich source of information on stellar evolution theories, allowing us to determine parameters such as the mass, temperature, or absolute luminosity of their components. Eclipsing binaries, in particular, enable sensitive tests of stellar models, and are among the most accurate distance indicators \citep[e.g.][]{thompson01, pietrzynski13}.

Although variability surveys have been undertaken since the late nineteenth century, the advent of modern charge-coupled device (CCD) cameras ushered a new era in this type of studies, enabling us to monitor millions of stars regularly, and to detect even low-amplitude variability. Microlensing surveys, for instance the Massive Compact Halo Object project (MACHO, \citealt{alcock93}), Exp\'erience de Recherche d'Objets Sombres (EROS experiment, \citealt{aubourg93}) and Optical Gravitational Lens Experiment (OGLE, \citealt{udalski92}), have revolutionized our knowledge of these objects in fields towards the Galactic Bulge and the Magellanic Clouds. In particular, OGLE observations led to the detection of many tens of thousands of pulsating variable stars, including some new types \citep{pietrukowicz17}, as well as hundreds of thousands of binaries \citep{soszynski16}. Other significant variability surveys leading to large numbers of new variables being detected include the All Sky Automated Survey (ASAS, \citealt{pojmanski02}), the Catalina Sky Survey (CSS, \citealt{drake14}), the Panoramic Survey Telescope and Rapid Response System (PanSTARRS, \citealt{hernitschek16}) survey, Palomar Transient Factory (PTF, \citealt{rau2009}), the VISTA Variables in the V\'{\i}a L\'actea (VVV, \citealt{minniti10}). Moreover, work on variability detection algorithms have also led to thousands of additional detections on already existing data \citep[e.g.][]{devor05}. 

In this paper we present light curves for the 320 new detected variables and candidate variables, based on observations obtained with the VIMOS camera at the Very Large Telescope (VLT, ESO). These observations were taken over the years 2013 and 2014, with the primary aim of providing ground-based follow-up of microlensing events \citep{kains17}.
However, the observing cadence of $\sim$ 4 days and the long baseline make this data set an excellent resource for the detection of variable stars. The depth of the survey afforded by the use of the VLT enabled us to detect new variables in field that were already observed extensively by OGLE. 
In Section \ref{sec:observations}, we detail our observations, before discussing our data reduction pipeline, photometric calibration and astrometry in \Sec{sec:reduction}. We describe the techniques used to identify variables in \Sec{sec:varsearch}, and analyze and classify the light curves in \ref{sec:lightcurves}. The final catalog is outlined and discussed in \Sec{sec:discussion} and conclusions are illustrated in \Sec{sec:conclusions}.

\section{Observations}\label{sec:observations}

We observed three pointings towards the Galactic Bulge with VIMOS on the VLT as part of a program\footnote{ESO-091.D-0489(A) and 093.D-0522(A), PI: M. Zoccali} designed to monitor Bulge stars simultaneously with the \textit{Hubble Space Telescope} (\textit{HST})\footnote{HST GO-120586, GO-13057, and GO-13464, PI: K.C. Sahu} and from the ground. These observations were taken in order to constrain the effect of the Earth's orbit around the Sun on microlensing event light curves, aiming to determine distances to lenses in these events. Combined with the \textit{HST} observations, such measurements can enable direct mass measurement of potential isolated stellar-mass black holes, through the detection of both photometric and astrometric microlensing \citep{kains17}.

\begin{figure}
\begin{center}
\includegraphics[height=0.33\textheight,width=0.45\textwidth]{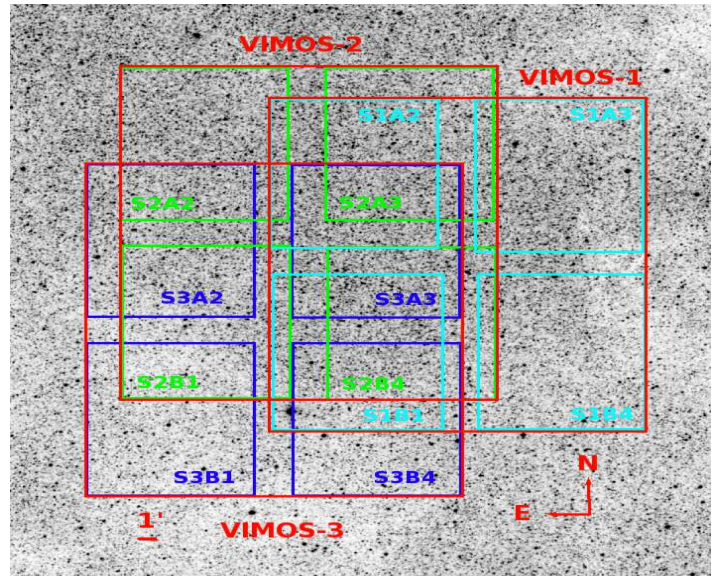}  
\caption{Finding chart for the three VIMOS pointings observed towards the Galactic Bulge SWEEPS window. The four quadrants of each VIMOS field are also shown and labeled. Coordinates for each pointing are listed in \Tab{tab:vimosobs}.\label{fig:fov}}
\end{center}
\end{figure}

The three VIMOS pointings are listed in \Tab{tab:vimosobs}, and shown in the finding chart in \Fig{fig:fov} with red boxes. VIMOS is a wide-field imager composed of four different arms, each with a field of view of $7\arcmin$ by $8\arcmin$. The detector pixel scale is $0.205\arcsec$, and the gap between each quadrant is $\approx 2\arcmin$ (see \Fig{fig:fov}). Because there is significant overlap between the three pointings, the total effective covered area is $\sim450$ arcmin$^2$. For further details on this instrument, we refer the reader to \citet{lefevre03} and  \citet{hammersley10}. The VIMOS observations were obtained in 2013 and 2014, from early April to early October, at a cadence of approximately one epoch every 4 days, which satisfied the program's primary science goal of monitoring microlensing events with coverage sufficient to detect the effect of the Earth's orbit on microlensing light curves. Images were obtained in the Bessel-$I$ filter, and some Bessel-$V$ band images were also taken during the 2014 season. Exposure times were 30s for most images, and a smaller number of 10s exposures were collected in order to construct a stacked reference image with fewer saturated stars in the field of view. \Tab{tab:vimosobs} summarizes the number of images available for each filter. The observing conditions were quite good on average, with seeing ranging from 0.38$\arcsec$ to 1.74$\arcsec$ during the first year, and from 0.43$\arcsec$ to 2.26$\arcsec$ in the second year of observations.

%% =====================================================
\begin{table}
  \vspace{0.5cm}
\begin{center}
  \begin{tabular}{ccccc}
\hline
Pointing	&RA	&Dec	&$I$ 	&$V$  \\
\hline	   
VIMOS-1	&17:58:42.9	&-29:15:14	&815		&9\\
VIMOS-2	&17:59:11.8	&-29:13:33	&811		&6\\
VIMOS-3	&17:59:18.2	&-29:18:56	&818		&6\\
\hline
  \end{tabular}
  \caption{Summary of our VLT/VIMOS observations, with the number of epochs in $I$ and $V$ for each pointing. \label{tab:vimosobs}}
  \end{center}
\end{table}
%% =====================================================

\section{Data reduction}\label{sec:reduction}
After downloading the raw images and their associated calibration files from the ESO archive, we applied bias and flat-field corrections to each image, by using biases and twilight sky flat field images collected during the same observing night or the closest set to the night\footnote{Flat-field images were  taken typically within 7 nights.}. All the exposures for each quadrant are then analysed separately. 

\subsection{Difference Imaging}
We performed photometry on the VIMOS images using the difference image analysis (DIA) pipeline {\tt DanDIA} \citep{bramich13,bramich12b, bramich08}, which works particularly well in crowded fields such as the cores of globular clusters \citep[e.g.][]{kains15a,kains13b, kains12b,figuera13,bramich11} and the Galactic Bulge \citep[e.g.][]{kains13a}. Each VIMOS quadrant was reduced separately with its own reference image (see \Fig{fig:fov} for the quadrant naming and distribution).

The reference image was produced by stacking the short (10s) exposures taken under seeing conditions within 10\% of the best-seeing images. The shorter exposures minimize the number of saturated stars that are present in the longer-exposure (30s) images. This resulted in a reference image with an effective exposure times of 60-180 seconds, and with a point source Full-Width Half Maximum of 0.6-0.7$\arcsec$. For more details on the data reduction technique used we refer the reader to the paper of \cite{kains17}, who used the same data set and reduction to analyse microlensing events. 

\begin{figure}
\begin{center}
\includegraphics[width=8cm, angle=0]{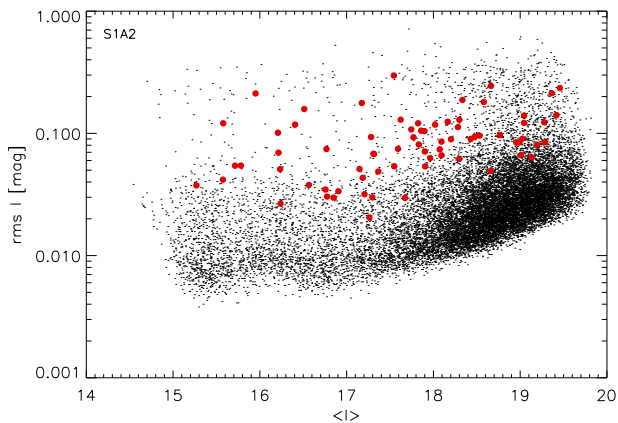}
\caption{$I-$band rms plot for one of the quadrants. All stars in the pointing are plotted as black dots, with variables (both previously known and new) that are discussed further in the text plotted as red filled circles. Many of the large-rms (>0.1 mag) stars are close to bright or saturated stars, rather than real variables. \label{fig:rms}}
\end{center}
\end{figure}

The photometric precision, quantified by the root mean square scatter of our final light curves, can be seen in \Fig{fig:rms}. The plot shows that we reach a limiting magnitude of $I \approx$ 19.7 with a signal to noise ratio $S/N \ge$ 20.

Photometric catalogues with average magnitudes in the $I$ and $V$ filters for all stars in each of the 12 quadrants were produced. These contain $\sim$40,000 stars per catalogue, a total of almost half a million stars in the entire field of view.

\subsection{Photometric calibration}
The photometric calibration was performed by using the OGLE-II $VI$-band catalog of the Galactic bulge \citep{udalski02} as a set of secondary standard stars. The OGLE filter system closely resembles the standard system: the $I$ filter is the Kron-Cousins $I$, while the transmission of the OGLE $V$ filter is slightly different from the standard Johnson $V$ \citep{udalski2015}.
The DIA analysis produced a $VI$-band catalog for each quadrant with an average photometric accuracy of $\sigma_I \approx$ 0.01 mag and $\sigma_{(V-I)} \le$ 0.04 mag at the turn-off level, $I \approx$ 18.7 mag. The photometric error is dominated by the $V$-band photometry, for which we only have a total of 9 exposures per field, all taken in the second season of observations.
We matched each of the 12 catalogs to OGLE photometry and selected stars with best photometric accuracy in both catalogs ($\sigma (V,I) \le$ 0.025 mag and number of exposures, $n_{V} \ge$ 9 and $n_{I} \ge$ 250), ending up with $\sim$ 150-200 stars per field. The calibration equation was derived as:

\begin{equation}
   V,I_{\rm cal} = V,I_{\rm inst} + \alpha + \beta \times (V-I)_{\rm inst}
	\label{eq:vical}
\end{equation}

where $\alpha$ and $\beta$ are the measured constant and linear coefficient, respectively, and $V,I_{\rm inst}$ and $(V-I)_{\rm inst}$ are the instrumental magnitudes and colors.
The coefficients of the calibration curves and their errors for each VIMOS quadrant are listed in Table~\ref{tab:curves}. The typical errors range between 1\% to 3\%.

\begin{figure}
\begin{center}
\includegraphics[height=0.55\textheight,width=0.45\textwidth]{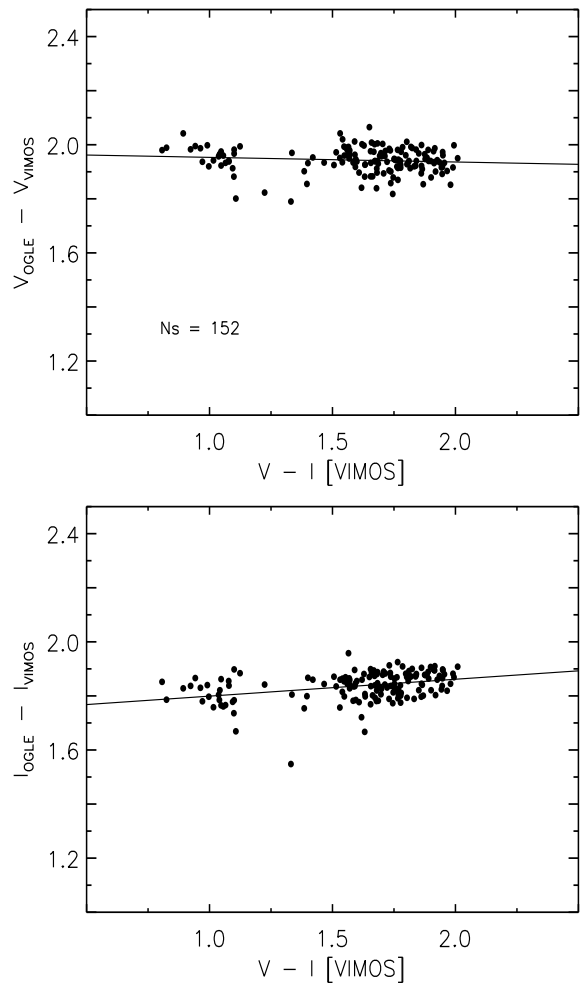}  
\caption{Calibration equation fits for the $V$ (top panel) and the $I$ (bottom) filter based on the comparison of VIMOS and OGLE photometry of 152 stars in common in field S1A2.\label{fig:cali}}
\end{center}
\end{figure}

\Fig{fig:cali} shows the calibration curves for the $V$ (top panel) and the $I$ (bottom) filters for field S1A2, based on the photometry of 152 common stars with OGLE. 
S1A2 was selected as a reference field because of its largest overlap with the ACS SWEEPS field (Sagittarius Window Eclipsing Extrasolar Planet Search; \citealt{sahu06, clarkson08}), and it is one of the field the least affected by reddening. In order to derive the extinction towards the observed VIMOS fields, we used the online reddening beam calculator\footnote{\url{http://mill.astro.puc.cl/BEAM/calculator.php}} provided by \cite{gonzalez11, gonzalez12}, which is based on data from the VVV Survey \citep{minniti10} and the Two Micron All-Sky Survey (2MASS, \citealt{skrutskie06}). We used the reddening law of \cite{cardelli89} and estimated the average extinction in the middle of a box of side 8$\arcmin$ centered on each VIMOS field. $A_V$ and $E(B-V)$ were derived from $A_K$ as $A_V$ = $A_K/0.11$ and $E(B-V)$ = $A_V/3.1$. The reddening values for the 12 VIMOS quadrants range from $\approx$ 0.7 to 1 mag and are listed, together with uncertainties from the BEAM extinction calculator, in Table~\ref{tab:reddening}. 

As a consistency check for the derived extinctions, we also derived reddening values using the light curves of RR Lyrae in each of the fields. Indeed, as first proposed by \cite{sturch66}, and further improved by \cite{guldenschuh05} and by \cite{kunder10}, the color of RRab stars near minimum brightness can be used to derive a reddening value, given an intrinsic color. Using an intrinsic color $(V-I)_0=0.58\pm0.02$ mag between phases 0.5 and 0.8 \citep{guldenschuh05}, we derived the values of $E(V-I)_{\rm RRL}$ listed in Table~\ref{tab:reddening} with their errors. We find these to be consistent, within uncertainties, with the values derived from the VVV reddening estimates.

\begin{figure}
\begin{center}
\includegraphics[height=0.35\textheight,width=0.45\textwidth]{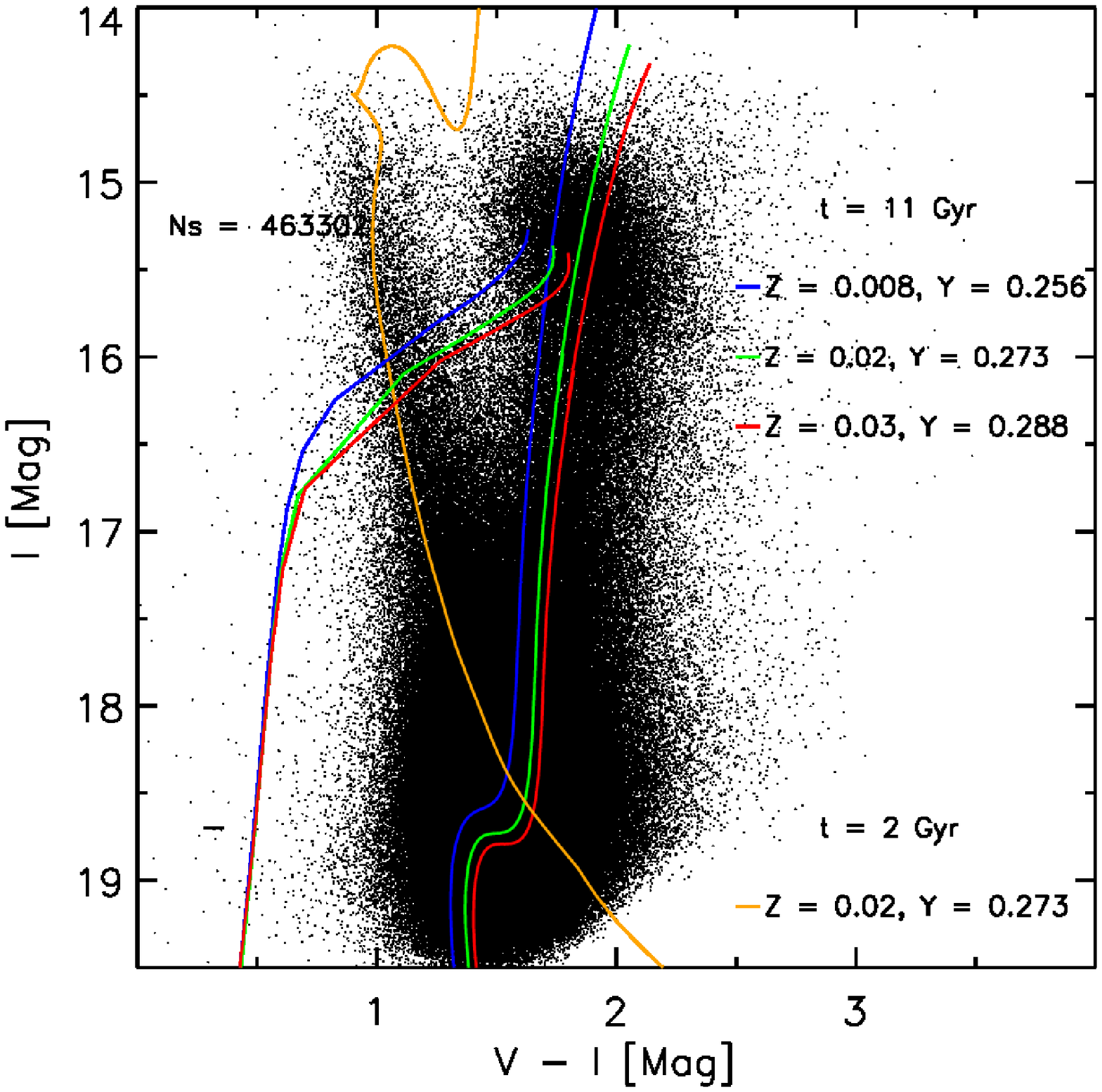}  
\caption{$I, V-I$ color-magnitude-diagram of all the stars observed in
the 12 VIMOS quadrants in 3 fields. Magnitudes and colors are corrected for the reddening differences between the quadrants but not for differential reddening. Isochrones for different ages and metallicities and ZAHBs are also shown. The metallicities, helium abundances and ages are labeled. 
Error bars indicate the average magnitude and color errors at the main-sequence turn-off point. \label{fig:cmd}}
\end{center}
\end{figure}

The photometric catalog of each VIMOS field was calibrated by using the calibration equations with coefficients listed in Table~\ref{tab:curves} and the photometry of all quadrants was shifted to the photometry of the reference field, S1A2, by applying the derived differences in reddening. We then obtained a final merged catalog including $\sim$ 460,000 stars. Note that we do not correct magnitudes for exinction and differential reddening as it is not necessary for the purposes of this study. \Fig{fig:cmd} shows the $I, V-I$ color-magnitude diagram (CMD) based on the final calibrated catalog. The catalog includes both bulge and disk stars: a blue plume extending from $I \approx$ 14.5 down to $\approx$ 17.5 mag is clearly visible in the CMD, an extended red-clump is present in the magnitude range 15 $\lesssim I \lesssim$ 16.5 and color 1.5 $\lesssim V-I \lesssim$ 2.5. Due to the limit of photometric accuracy, stellar crowding and the presence of differential reddening, it is not possible to identify the main-sequence Turn-Off (MSTO) in our CMD. Error bars indicating the average magnitude ($I$) and color errors at the level of the MSTO are shown in \Fig{fig:cmd}.

To validate the current photometric calibration we used theoretical models. Isochrones for different metallicities and ages were retrieved from the BASTI database\footnote{\url{http://albione.oa-teramo.inaf.it/}} and over-plotted to the $I, V-I$ CMD in \Fig{fig:cmd}. To fit the Galactic bulge stellar populations we used scaled-solar models with the same age, $t =$ 11 Gyr, and different metallicities, ranging from $Z = 0.008$ to $Z = 0.03$, and a distance modulus of $\mu_0 =$ 14.50 mag and reddening $E(B-V)=$ 0.5 mag, following the prescriptions of \citet{calamida14b}. A younger 2 Gyr solar metallicity isochrone (yellow line) is also over-plotted to reproduce the average Galactic disk stellar populations. \Fig{fig:cmd} shows that, within uncertainties, theory and observations are in good agreement.

%% =====================================================
\begin{table*}
  \vspace{0.5cm}
\begin{center}
  \begin{tabular}{lcccc}
\hline
Field & $\alpha_{V}$ & $\beta_{V}$ &  $\alpha_{I}$ &  $\beta_{I}$ \\
\hline	   
S1A2 & 1.97 $\pm$ 0.02 & -0.017 $\pm$ 0.01  & 1.737 $\pm$ 0.02 & 0.062 $\pm$ 0.01\\
S1A3 & 1.92 $\pm$ 0.02 & -0.024 $\pm$ 0.01  & 1.756 $\pm$ 0.02 & 0.055 $\pm$ 0.01 \\
S1B1 & 1.89 $\pm$ 0.02 & -0.015 $\pm$ 0.01  & 1.591 $\pm$ 0.02 & 0.072 $\pm$ 0.01 \\
S1B4 & 1.94 $\pm$ 0.04 & -0.010 $\pm$ 0.02  & 1.678 $\pm$ 0.03 & 0.072 $\pm$ 0.016\\
S2A2 & 1.94 $\pm$ 0.03 & -0.045 $\pm$ 0.02  & 1.721 $\pm$ 0.02 & 0.049 $\pm$ 0.014 \\
S2A3 & 1.81 $\pm$ 0.01 & -0.019 $\pm$ 0.01  & 1.681 $\pm$ 0.01 & 0.051 $\pm$ 0.006\\
S2B1 & 1.84 $\pm$ 0.01 & -0.026 $\pm$ 0.005 & 1.670 $\pm$ 0.01 & 0.049 $\pm$ 0.005 \\
S2B4 & 1.81 $\pm$ 0.02 & -0.032 $\pm$ 0.01  & 1.686 $\pm$ 0.02 & 0.057 $\pm$ 0.01 \\
S3A2 & 1.55 $\pm$ 0.02 & -0.018 $\pm$ 0.01  & 1.725 $\pm$ 0.02 & 0.043 $\pm$ 0.01 \\
S3A3 & 0.77 $\pm$ 0.02 & -0.026 $\pm$ 0.006 & 1.642 $\pm$ 0.02 & 0.052 $\pm$ 0.006\\
S3B1 & 1.55 $\pm$ 0.03 & -0.043 $\pm$ 0.01  & 1.724 $\pm$ 0.03 & 0.038 $\pm$ 0.01 \\
S3B4 & 0.81 $\pm$ 0.05 & -0.052 $\pm$ 0.02  & 1.673 $\pm$ 0.05 & 0.042 $\pm$ 0.02 \\
\hline
\end{tabular}
  \caption{Coefficients of the calibration curves for the 12 VIMOS quadrants in 3 fields.\label{tab:curves}}
  \end{center}
\end{table*}

%% =====================================================
\begin{table*}
  \vspace{0.5cm}
\begin{center}
  \begin{tabular}{lccccc}
\hline
Field	&  l   &   b   & $A_V$ 	& $E(V-I)$ & $E(V-I)_{\rm RRL}$ \\
\hline	   
s1a2  & 1.30 & -2.66 & 1.702 &  0.65$\pm$0.10 & $0.86 \pm 0.09$ \\
s1a3  & 1.20 & -2.52 & 2.033 &  0.77$\pm$0.11 & $0.92^1$ \\
s1b1  & 1.15 & -2.73 & 1.952 &  0.74$\pm$0.11 & $0.81^1$ \\
s1b4  & 1.08 & -2.59 & 2.379 &  0.90$\pm$0.11 & $1.12 \pm 0.02$ \\
s2a2  & 1.37 & -2.74 & 1.728 &  0.66$\pm$0.10 & $0.77^1$ \\
s2a3  & 1.28 & -2.59 & 1.734 &  0.66$\pm$0.10 & $0.84 \pm 0.10$ \\
s2b1  & 1.24 & -2.82 & 2.200 &  0.84$\pm$0.10 & - \\
s2b4  & 1.19 & -2.67 & 1.767 &  0.67$\pm$0.10 & - \\
s3a2  & 1.30 & -2.81 & 2.016 &  0.77$\pm$0.10 & $0.82 \pm 0.09$ \\
s3a3  & 1.22 & -2.66 & 1.736 &  0.66$\pm$0.10 & - \\
s3b1  & 1.17 & -2.88 & 2.383 &  0.91$\pm$0.11 & $1.01 \pm 0.02$ \\
s3b4  & 1.09 & -2.74 & 2.276 &  0.86$\pm$0.11 & $1.11 \pm 0.31$ \\
\hline
  \end{tabular}
  \caption{Average reddening values derived for the 12 VIMOS quadrants in 3 fields. Reddening values derived from a single RR Lyrae light curve are marked with $^1$ in the last column.\label{tab:reddening}}
  \end{center}
\end{table*}
%% =====================================================

\subsection{Astrometry}\label{sec:astrometry}

We used the ESO software {\tt Skycat}\footnote{\url{http://archive.eso.org/cms/tools-documentation/skycat.html}} and the USNO-B catalogue of astrometric standards \citep{monet03} to correct the astrometric solution for each of the 12 VIMOS reference images. The large distortion of the images results in the astrometric solution being very good (with residual scatter of $\sim0.2\arcsec$) near the central $\sim$ 500 $\times$ 500 pixels of each image, while being slightly off the true position near the edges by as much as 2$\arcsec$. However, this level of accuracy is sufficient for the purposes of identifying variable stars and cross-matching them with catalogs of known variables published by surveys like OGLE, ASAS, or VVV.

\section{Variability search}\label{sec:varsearch}
\subsection{Identifying variables}

We carried out searches for variable stars by using a number of independent methods. We first used the variability index $S_R$, as defined by \cite{kains13a,kains15a}; briefly, this quantifies the improvement in the phased light curve using the best-fit period compared to a random period. This is done by measuring the respective string lengths of the phased light curves, i.e. summing the distance between each successive pair of neighbouring data points in the phased light curve for each trial period, and identifying the period that minimises the total string length. We imposed a cut at $S_R=0.3$ for a variable to be retained in our sample. We also identified stars with large root mean square (rms) variation from the rms diagram (Fig. \ref{fig:rms}). To do this, we grouped stars into mean magnitude bins, and inspected light curves of stars with scatter above the 3-$\sigma$ level for each magnitude bin. Finally, we also stacked the absolute values of difference images, producing an image on which variables stand out because of their residual differential flux.

The combined methods resulted in a sample of 1143 periodic variables, most of which were found with at least two of the variable identification methods.

\subsection{Cross-matching with known variables}

We cross-matched our sample of variables with known variable catalogues, including those of \cite{devor05}, OGLE-III \citep{udalski08,soszynski11a,soszynski11b,soszynski13}, the OGLE catalogue of eclipsing and ellipsoidal variables in the Galactic Bulge \citep{soszynski16}, ASAS \citep{pojmanski02}, the \textit{Gaia} \citep{gaia16} data release 2 (DR2) catalogue \citep{gaiadr2}, and all variables listed by the SIMBAD astronomical database \citep{wenger00} within our VIMOS pointing. Given the uncertainties on our astrometry, particularly away from the centre of the images, we used a matching radius of 3$\arcsec$ to ensure that we did not miss matches to known variables and claim erroneous new detections. 
As expected, given significant overlap between our fields, we found that many of our variables were previously detected by OGLE; however, we identified 320 previously uncatalogued variables.

In order to assess the completeness of our sample, we compared the number of detected known OGLE variables in our VIMOS footprints with the total number of OGLE variables in our pointings. We find that all RR Lyrae stars detected by OGLE are recovered, but that several eclipsing and ellipsoidal variables have a variability index that is higher than our cut. However, many of these are present in our data, and after recovering those, we found an overall detection rate for known variables, of $\sim83\%$. The remaining $\sim 250$ OGLE variables that we did not detect mostly include stars with low amplitudes ($A_I \lesssim 0.05$ mag). In addition, they are distributed evenly across the CCDs, and across magnitude bins, meaning we did not miss them because of edge effects or because they were too faint; instead, many of these stars are close to saturated stars, which affected their photometry, and our ability to detect them.

\section{Light curve analysis}\label{sec:lightcurves}

\subsection{Period searches}
We determined periods for each light curve using the string length method that was used to determine the variability index $S_R$ mentioned in the previous section. This method can sometimes yield periods that are harmonics of the correct period when there is significant scatter in the light curve, and we therefore inspected all light curves visually to check whether this was the case. This was necessary for several cases where eclipses of slightly different depths were detected.

\subsection{Fourier decomposition and Principal component analysis}

We used the periods, determined in the previous section for each star, to fold the time-series data into a full-phased light curve. We performed Fourier decomposition \citep{simon81, bhardwaj15} and principal component analysis \citep[PCA,][]{deb2009} to obtain a preliminary classification of unknown variables. A Fourier sine-series was fitted to each light curve, after recursively removing $3\sigma$ outliers, in the following form:

\begin{equation}
m = m_{0}+\sum_{k=1}^{N}A_{k} \sin(2 \pi k x + \phi_{k}),
\label{eq:foufit1}
\end{equation}

\noindent where $m$ is the observed magnitude and $x$ is the phase of the light curve. The optimum order of fit ($N$) is obtained using the Baart's criteria \citep{baart1982,peterson1986} by varying it between 5-10. 
The amplitude and phase coefficients ($A_1,A_2\dots$ and $\phi_1,\phi_2\dots$) are used to formulate amplitude ratios ($R_{k1}=A_k/A_1$) and phase differences ($\phi_{k1}=\phi_k - k\phi_1$). These parameters have been often employed to study the light curve structure of variable stars \citep[for example,][]{bhardwaj15, kains12a}. The detailed mathematical formalism of PCA can be found in \cite{bhardwaj17}. We interpolated phased light-curves using cubic-spline fit to nearest neighbors and obtained 100 equally spaced data points for each light curve and used this to perform ensemble PCA analysis. In general, the lower-order Fourier coefficients or first few principal components (PC$_1$,PC$_2\dots$) can independently reproduce the light curve structure. 

\begin{figure*}
\begin{center}
\includegraphics[width=1.0\textwidth,keepaspectratio]{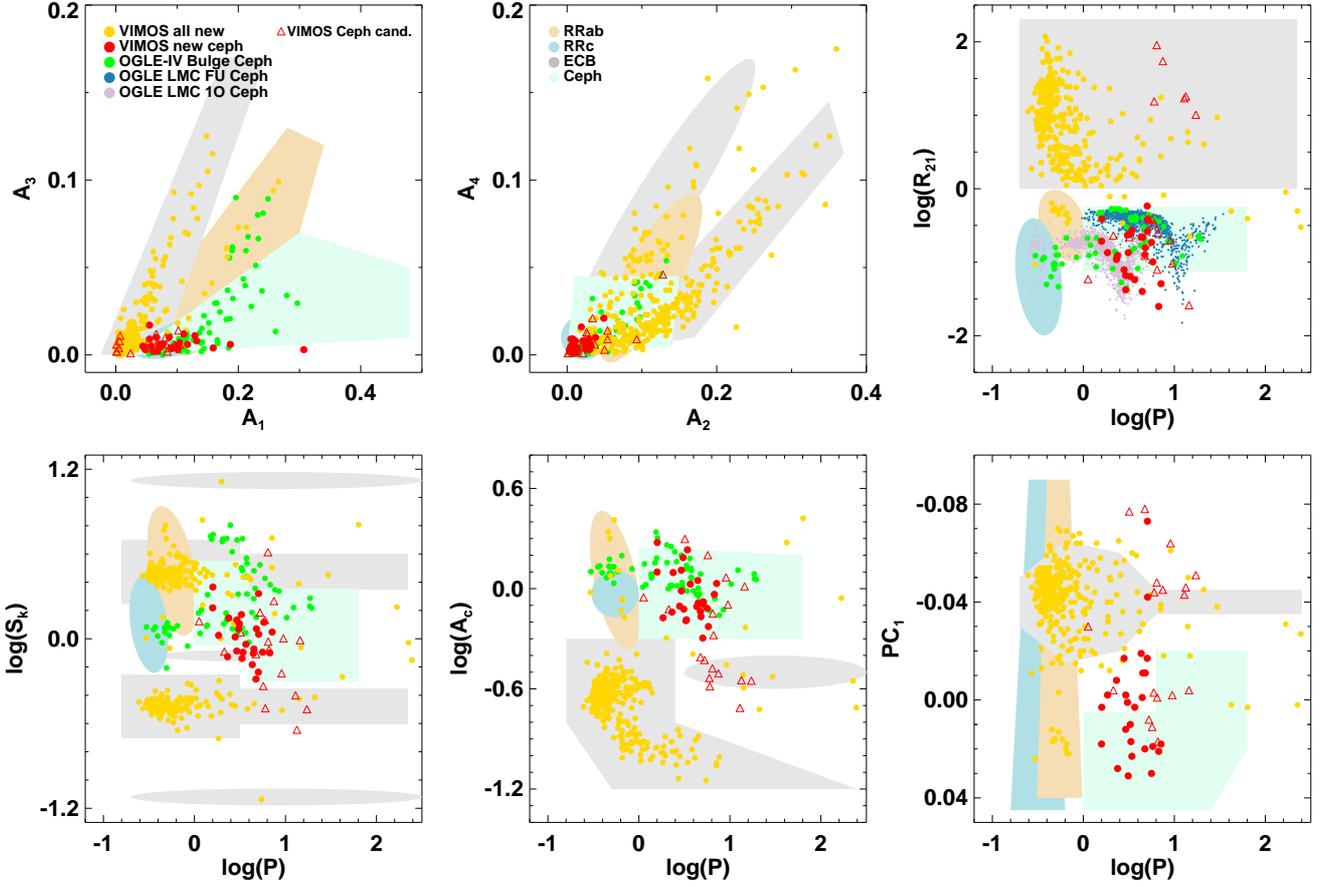}
\caption{The variation of different light curve parameters in $I$-band for the VIMOS variables. Fourier amplitude coefficients and amplitude ratio (top panels), skewness, acuteness parameters and first principal component (bottom panels) are used to display the separation of variables into different classes. Good Cepheid candidates identified in our sample (see Sec. \ref{sec:varclass}) are plotted as red filled circles. The approximate locations of these light curve parameters for the known variables in the Galactic Bulge from the OGLE survey \citep{soszynski11a,soszynski11b,soszynski14b,soszynski16} are also shown as shaded regions in the background. RRab and RRc denote fundamental and first-overtone RR Lyrae stars, respectively, while Ceph and ECB denote Cepheids and eclipsing binaries, respectively. Finally, individual Cepheids detected by OGLE in the Galactic Bulge, and in the LMC, are also shown as filled circles. Color codes for the individual stars are given in the top left panel, while legends for the shaded regions are shown in the top centre panel.}
\label{fig:lca}
\end{center}
\end{figure*}

Fig.~\ref{fig:lca} displays the variation of different light curve parameters of VIMOS variables for classification purposes. We also plotted the approximate locations of the known Bulge variables from the OGLE catalogue \citep{soszynski11a, soszynski14b, soszynski16} as shaded regions in the background. To identify the regions occupied by different types of variables, we selected the entire sample of RR Lyrae and Type II Cepheids, as well as $\sim 10,000$ representative eclipsing binary stars from the catalogue. In the top left and middle panels, selected lower-order Fourier coefficients are used to separate detached and contact eclipsing binaries \citep{pojmanski02, derekas2007}. The $A_1-A_3$ plot displays the separation between high-amplitude eclipsing and pulsating stars while $A_2-A_4$ plot shows the two distinct sequences of eclipsing binary stars typically belonging to the detached and contact systems \citep{derekas2007}. We note that the classification between contact binaries and first-overtone RR Lyrae stars is the most-uncertain as both are typically small-amplitude variables. The top right panel shows the most-distinct separation between various sub-classes of variable stars in the Fourier amplitude $R_{21}$ plane. The bottom left and middle panels display the variation in skewness and acuteness parameters as function of period. Skewness represents the left/right asymmetry while the acuteness displays the top/bottom asymmetry in the light curves. We note that eclipsing binary stars show different sequences of almost constant skewness values for the entire period range while some of the new variables fall in the shaded regions representing Cepheid stars. Similar features are also seen in the acuteness and first principal-component plane. We discuss our use of the automated classification algorithm of \cite{kim16} on our variables in the following section; this algorithm makes use of the PCA parameters to assign probabilities to different classes for each light curve.

\subsection{Categorising detected variables}\label{sec:varclass}

We used the classification algorithm {\tt Upsilon} \citep{kim16} to assign a class to each of our new variables, as well as an associated probability of the classification being correct. We divided the resulting classes into seven categories:
\begin{enumerate}
\item Contact binary: eclipsing and non-eclipsing contact variables, and semi-detached eclipsing binaries
\item Non-contact binary: detached eclipsing binaries
\item RR Lyrae stars
\item Cepheids: classical and type II
\item Delta Scuti, including both Delta Scuti and sub-types such as SX Phoenicis stars
\item Long-period variables: stars that do not fall in any of the above categories, but have clear periodic variations and periods longer than 20 days
\item Other: Star that do not fall in other categories
\end{enumerate}

In order to test the accuracy of the algorithm on our data set, we also used {\tt Upsilon} to classify the known variables in our sample, and compared it to the published classification. We find that {\tt Upsilon} performs well on most variable classes, but misclassifies many ellipsoidal variables and eclipsing binaries as RRc stars. However, the {\tt Upsilon} classification in the cases of these sources is characterised by a low probability score, $P_{\mathrm{class}} < 0.5$. In order to take this into account, we inspected light curves with low probability scores visually, in order to assess the accuracy of the classification from the algorithm.

{\tt Upsilon} classified many stars with periods of several days as contact binaries, which would imply very massive stars, and is unlikely to be correct. Upon visual inspection, we found instead that many of these stars are good Cepheid candidates. To test this, we fitted these stars with the Cepheid template light curves of \cite{inno15}, and examined the resulting fit to determine whether each of these star was a good Cepheid candidate, based both on the $\chi^2$ statistic of each fit, and on a visual inspection to assess whether the best fit matched the morphology of each light curve; this was necessary because of some objects having low-$\chi^2$ fits despite clearly not being well described by the best-fit model. In addition to this, we used the distribution of Fourier parameters (Fig. \ref{fig:lca}), compared to populations of known Cepheids in the Bulge and the Large Magellanic Cloud (LMC), to exclude some of the Cepheid candidate on the basis of their Fourier parameters. This resulted in a sample of 26 confirmed Cepheids with both good template fits and Fourier parameters consistent with those of known Cepheids, as well as 18 stars which satisfy only one of those criteria, and which we therefore classify as Cepheid candidates.
Finally, we also reclassified remaining Cepheids with periods longer than 20 days and low amplitudes as LPVs, as such low amplitudes are not consistent with populations of known Cepheids. Although DIA may underestimate the amplitude of variable stars in cases where significant blending is present, it is unlikely to be the explanation for most of the low-amplitude pulsators we detect here, further justifying our reclassification of these objects as LPVs.

\section{Final catalogue and discussion}\label{sec:discussion}

As expected, we found that most of our detected variables were already known, mainly in the extensive catalogue of OGLE \cite[e.g.][]{soszynski16}. However, thanks to  better spatial resolution afforded by the VLT, we also detected 320 variables that are not part of previously published catalogues.

The variable stars identified in VIMOS dataset were matched with the VVV 
Infrared Astrometric Catalogue \citep[VIRAC; ][]{smith18}, which contains highly
reliable proper motion measurements for over 119 million sources within the 
560 sq.deg VVV area, of which 47 million have statistical uncertainties below 1
mas~yr$^{-1}$. The relatively bright magnitude range of the VIMOS variables can be
matched well with the K-band VVV PSF-based photometric catalogues 
(Surot et al., in prep), however 88\% of our variables are fainter than $K_s =
14$, magnitude beyond which VIRAC proper motion errors increase quite rapidly
for fainter magnitudes. At a typical brightness of our sources, close to
$K_s=16$~mag, VIRAC proper motion errors are typically larger than 3 mas~yr$^{-1}$. 

To match the VIMOS variables catalogue with VIRAC sources we
downloaded VIRAC proper motions for the VVV tiles b292 and b306 from the public catalogue\footnote{\url{vvv.herts.ac.uk}}. We experimented with matching VIRAC and VIMOS catalogues with 1$\arcsec$ and 2$\arcsec$ maximum separation and verified the reliability of the matches by looking at the matched pairs distribution as well as the correlations between the optical VIMOS mean I magnitude and VVV mean $Z$, $Y$, $J$ and $K_s$ magnitudes for the matched sources. The VIRAC catalogues are on ICRS astrometric system with epoch 2012.0 and equinox 2000.0, which is very close in time to the epoch of our observations (2013-2014) and matching was thus quite straightforward with {\tt Topcat}. The match
with b292 tile yielded 761 matched variables within 2$\arcsec$, of which 440 are within 1$\arcsec$, and the match with b306, which has a much less overlap with VIMOS pointings, yielded 66 variables within 2$\arcsec$ (30 within 1$\arcsec$). The cross-match between VIMOS+b292 and VIMOS+b306 showed 5 groups where the same VIMOS variable was associated with two different VIRAC
sources (once in b292 and once in b306) in case of 2$\arcsec$ cross-match radius, but there were no duplicates for 1$\arcsec$ matches. The internal match of the 1143 VIMOS variables catalogue shows that 42 stars have $<1\arcsec$ distance. Increasing the internal match distance up to 2$\arcsec$, yields 68 matched groups. Given the crowded fields and the better correlation of magnitudes for the matched sources within 1$\arcsec$, to minimize spurious cross-matches we decided to keep only the matches made within 1$\arcsec$ radius.\footnote{Note that we could not use the matching magnitudes criteria to distinguish potentially spurious matches for all sources within 2$\arcsec$, because of long wavelength baseline and a range of possible $I-Ks$ colours, and lack of $Z$, $Y$ and $J$ measurements for some of the VIRAC sources.}
This means that out of 1143 VIMOS variables, we have proper motions from VIRAC for 470 stars, of which 211 are flagged as reliable in VIRAC, meaning that their proper motion have high quality, although due to faint sources their precision is rather low. Of the 320 new variables, 56 have reliable VVV proper motion measurements. The proper motions distribution of the VIMOS variables is consistent with a Bulge population \citep[e.g.\ ][]{clarkson08,vasquez13}, but large errors due to faint sources prevents a clear separation of bulge and disk stars. In the future, the VVV eXtended ESO Public Survey (VVVX), leading to VIRAC version II, should improve the proper motions \citep{smith18}.

In addition to the VIRAC proper motions, we also cross-matched our list of variables with the \textit{Gaia} DR2 catalogue \citep{gaiadr2}. We have retained both \textit{Gaia} and VIRAC proper motions when both were available, as VVV might perform better in crowded fields like the Galactic Bulge.

Parameters for our new variables can be found in the electronic version of this article. An excerpt from the table is shown in \Tab{tab:catalogue1} \& {tab:catalogue2} showing the content and format of the electronic data file. For each variable we provide coordinates, period, mean $I-$ and $V-$ band magnitudes from VIMOS as well as mean $J-$ and $K-$band magnitudes, proper motion values, and source ID from VIRAC, where available. 

\begin{figure*}
\begin{center}
\includegraphics[width=10cm,angle=90]{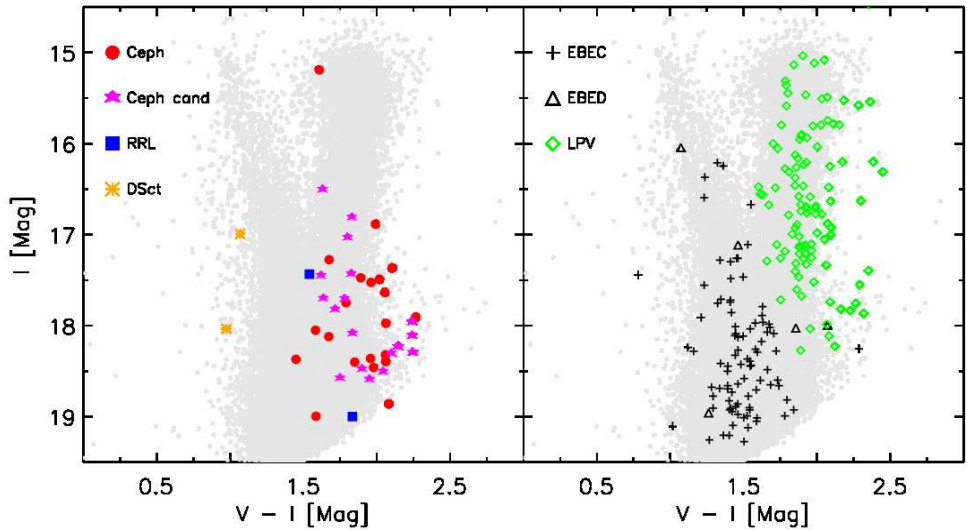}  \caption{Left: $I, V-I$ color-magnitude-diagram of stars observed in VIMOS field S1.A2 with the new identified RR Lyrae, Cepheids and Delta Scuti variables overplotted. Right: Same CMD with the new identified long period variables and contact and detached binaries. Different variable classes are marked with different colors and labeled in the figure. \label{fig:cmdvar}}
\end{center}
\end{figure*}

\begin{figure*}
\begin{center}
\includegraphics[width=7cm, angle=0]{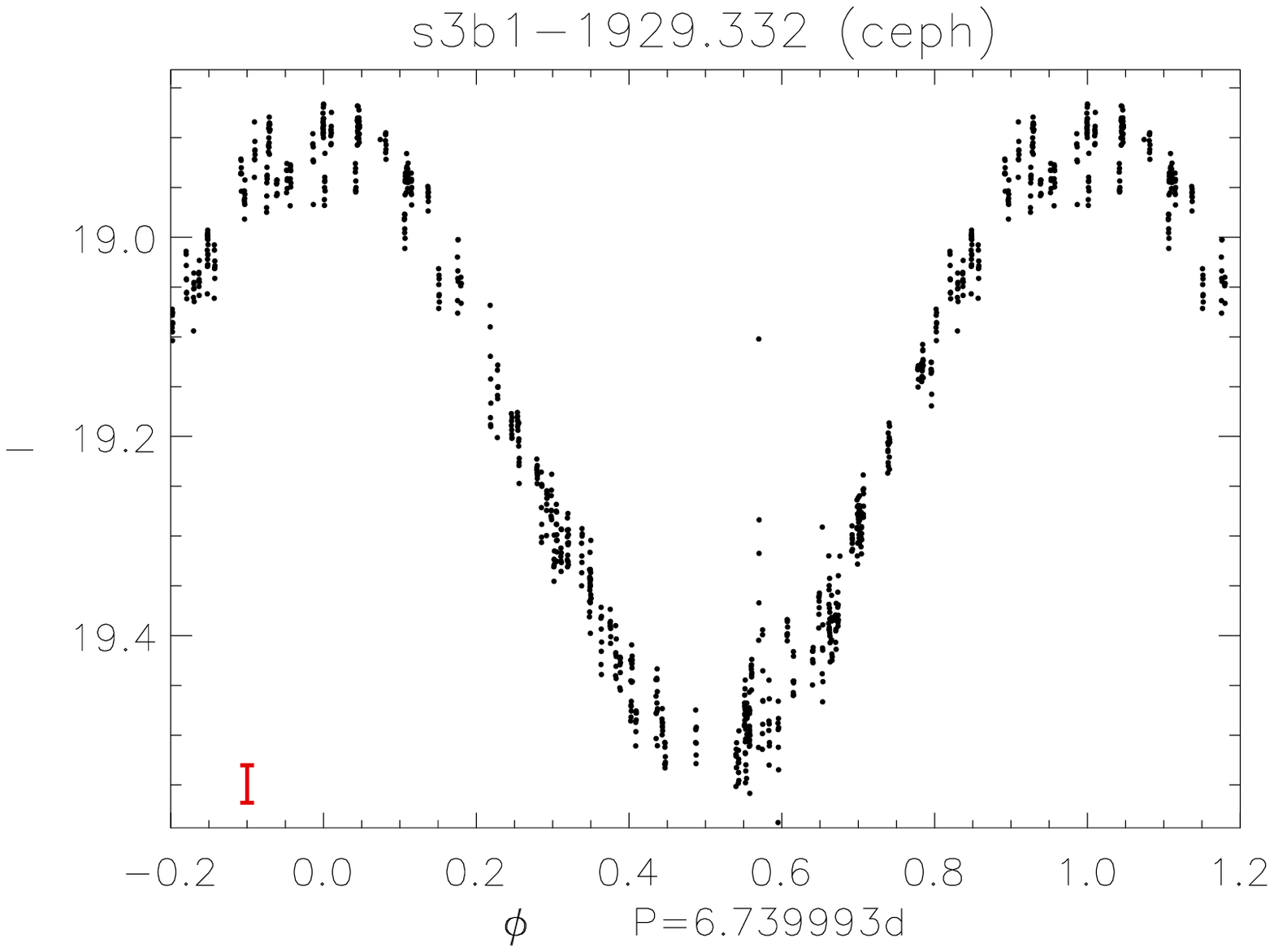}
\includegraphics[width=7cm, angle=0]{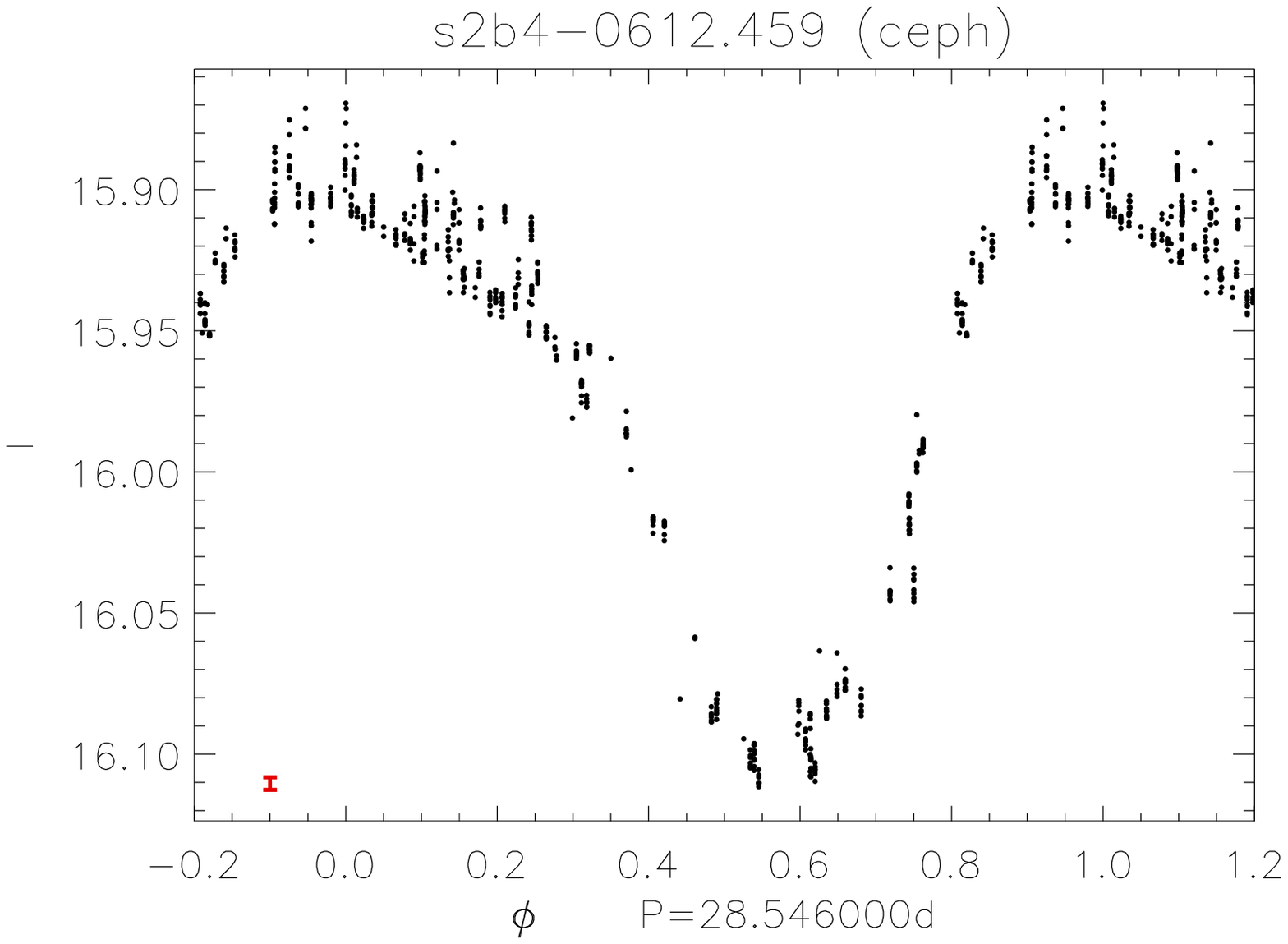}
\includegraphics[width=7cm, angle=0]{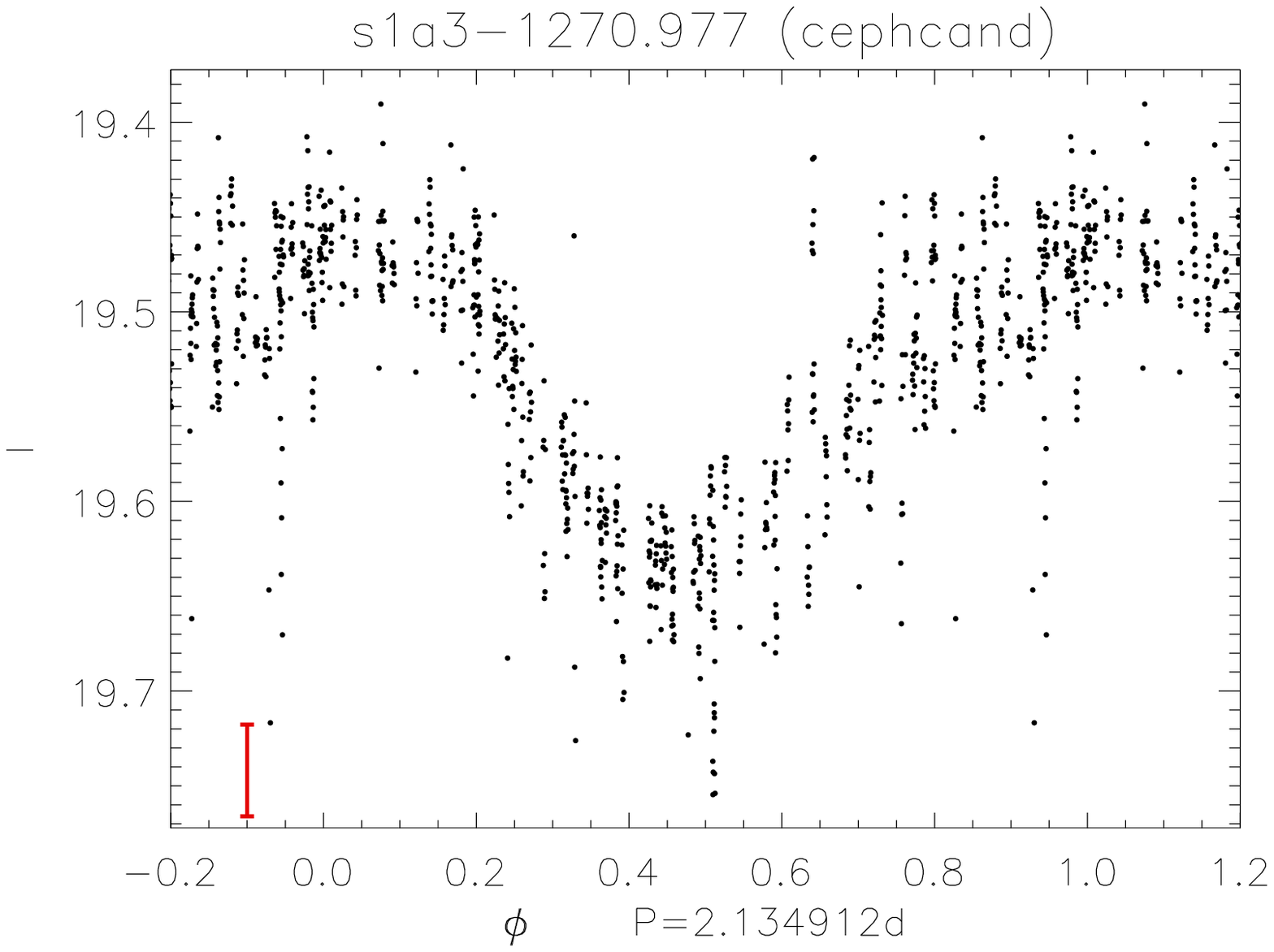}
\includegraphics[width=7cm, angle=0]{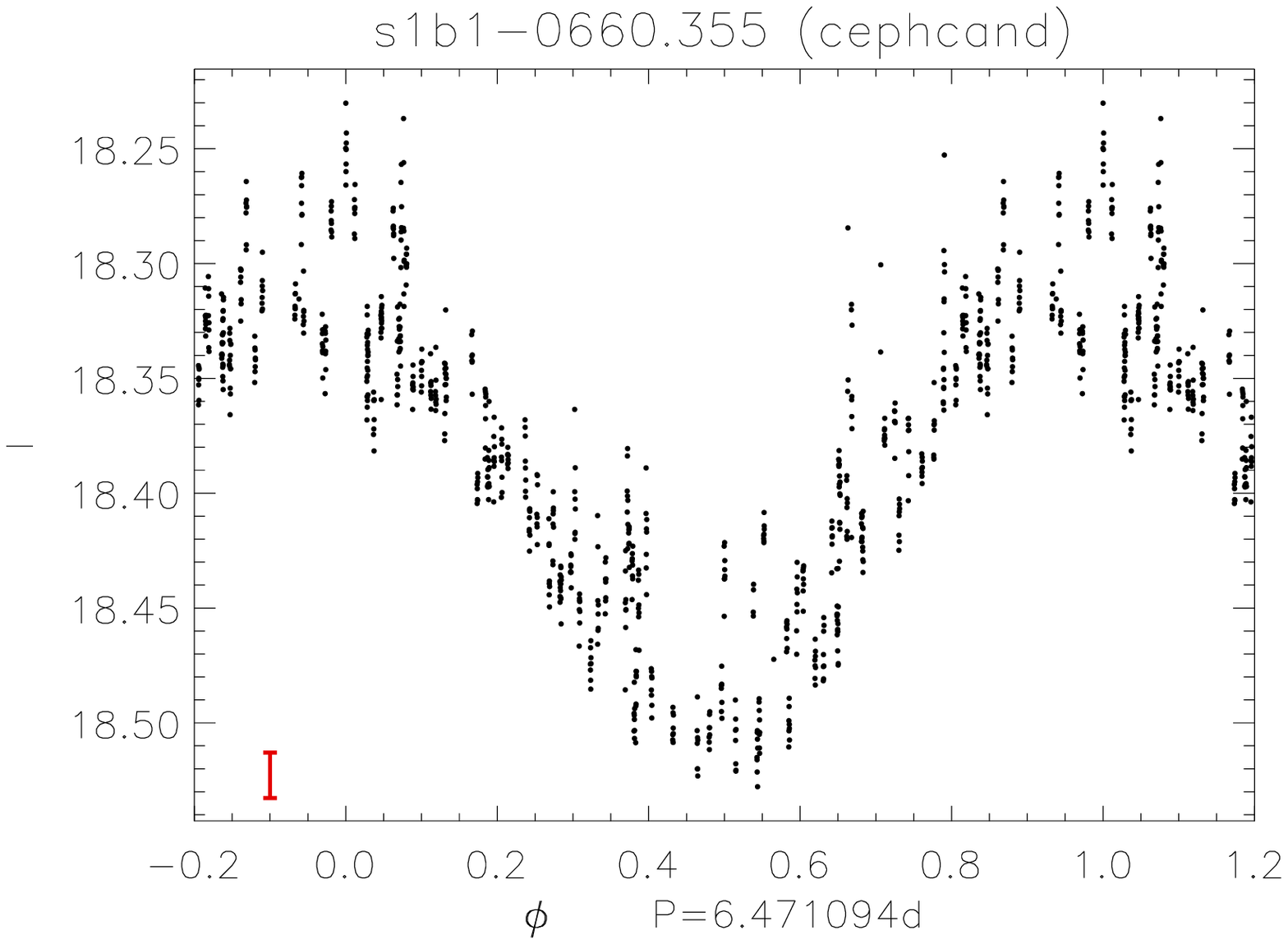}
\includegraphics[width=7cm, angle=0]{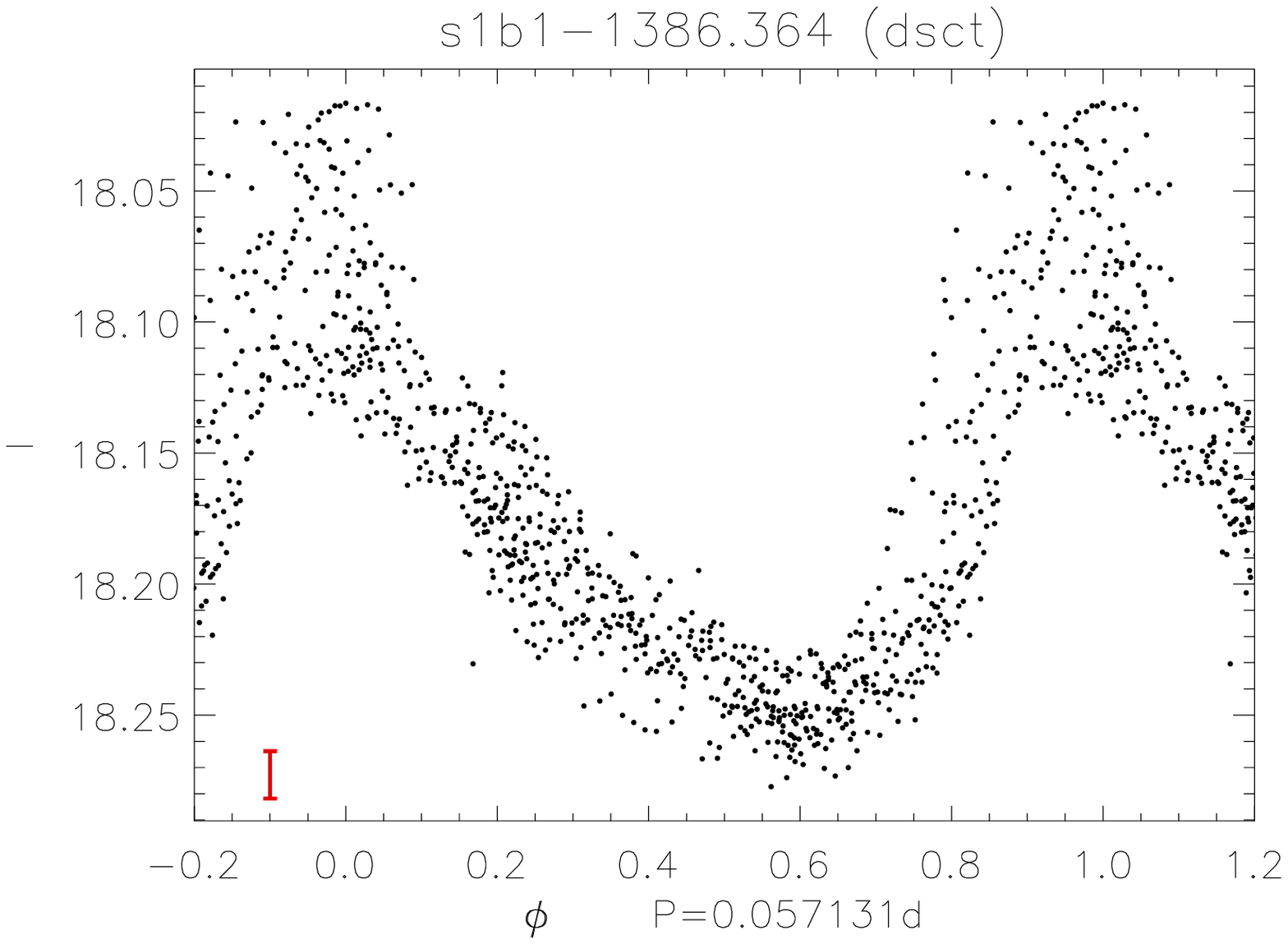}
\includegraphics[width=7cm, angle=0]{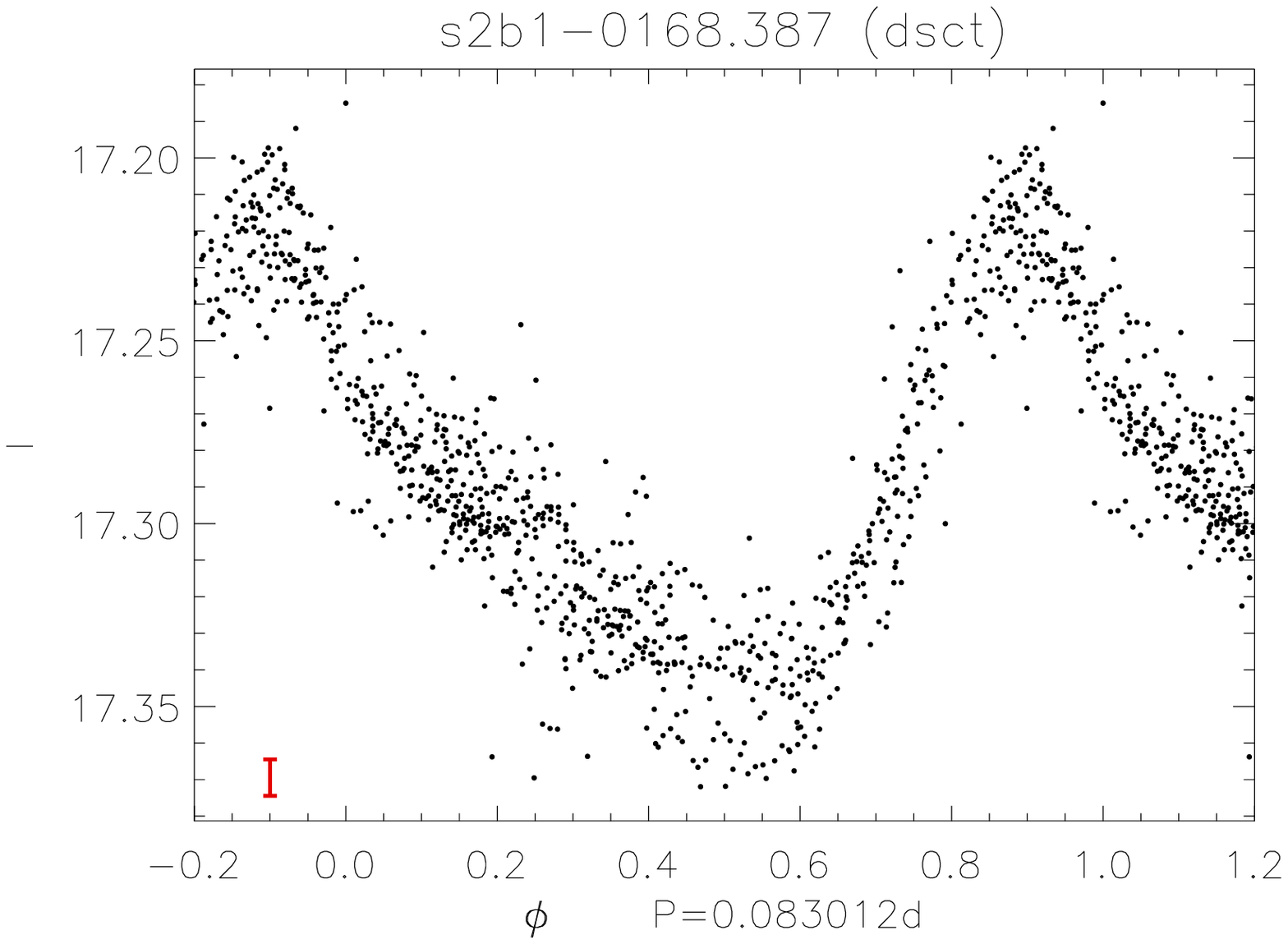}
\includegraphics[width=7cm, angle=0]{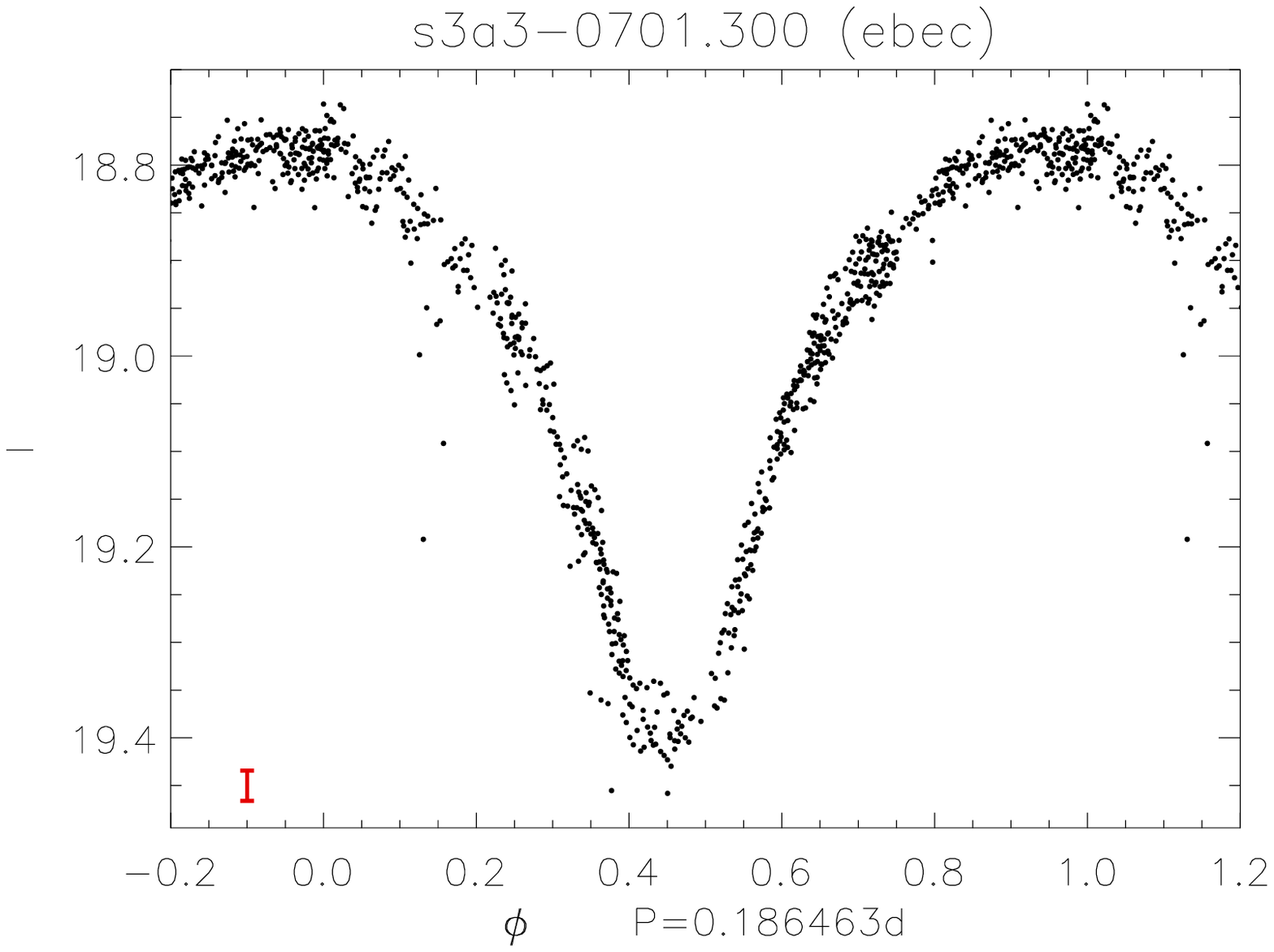}
\includegraphics[width=7cm, angle=0]{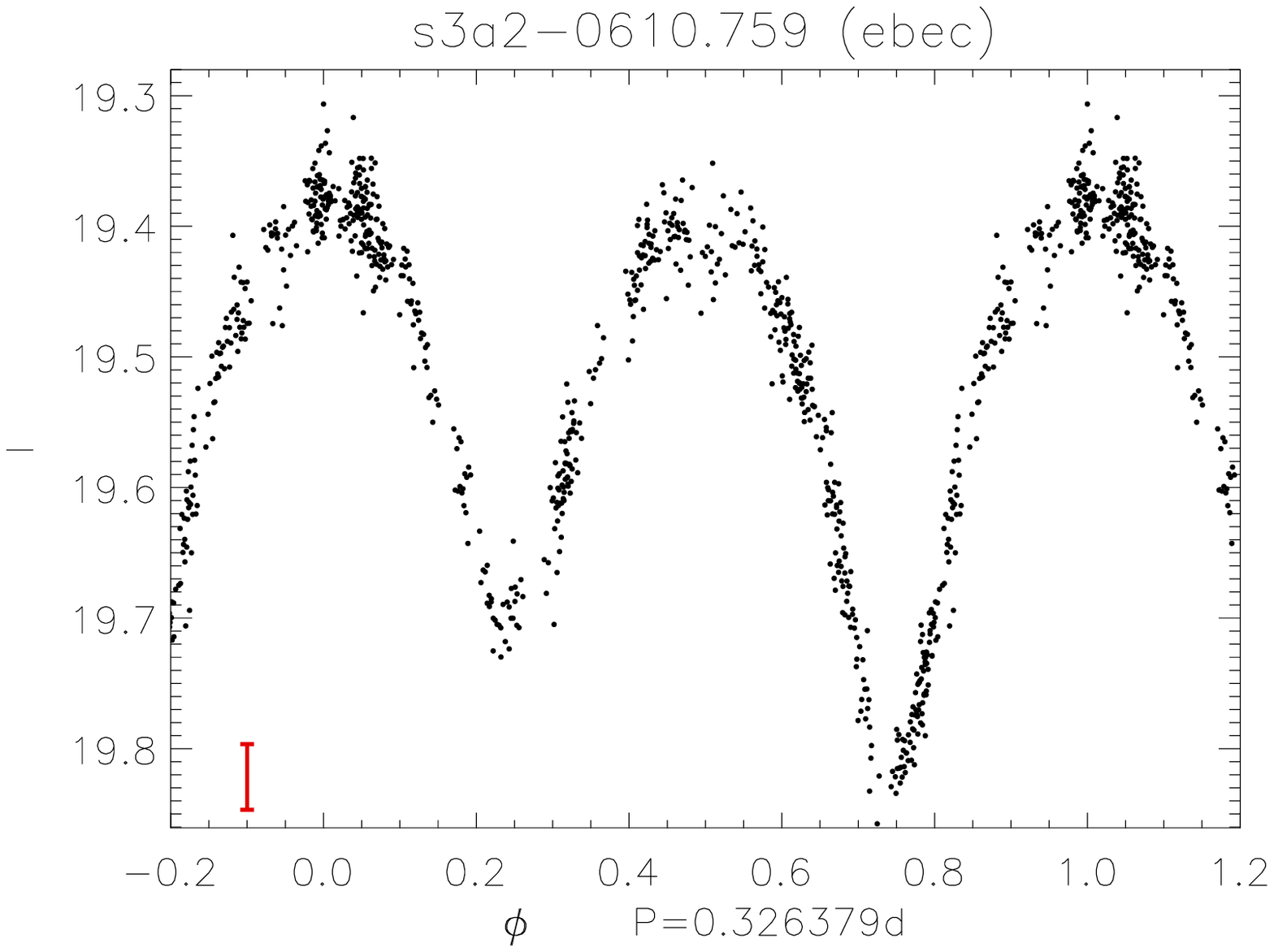}
caption{Sample phased light curve for each variable class. 2 representative light curves are shown for each class, which is specified in the title of each panel. For clarity, individual data points are plotted without error bars, but a typical error bar is plotted as a red bar in the bottom left of each panel. Continued in \Fig{fig:samplelc2}. \label{fig:samplelc1}}
\end{center}
\end{figure*}

\begin{figure*}
\begin{center}
\includegraphics[width=7cm, angle=0]{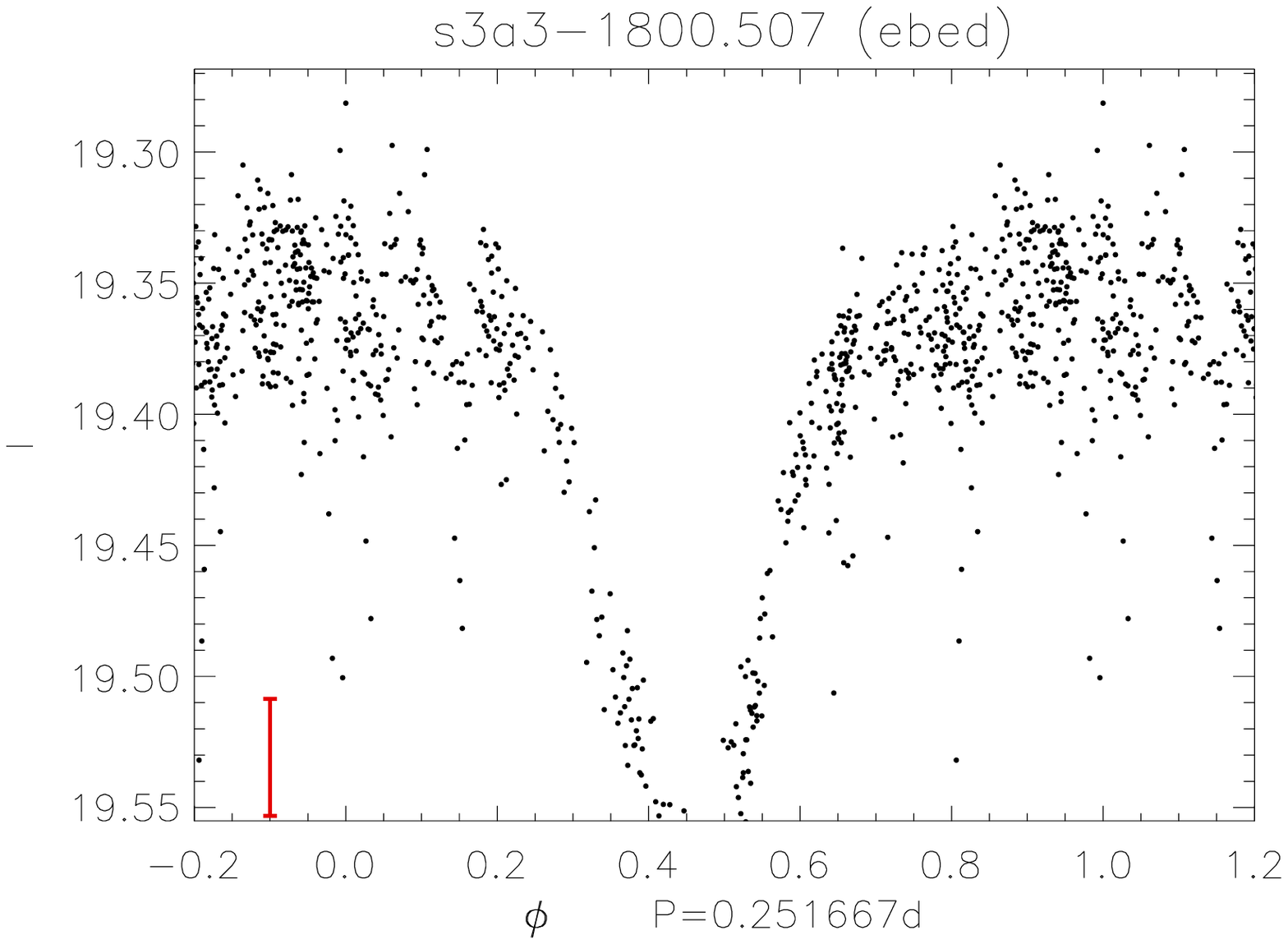}
\includegraphics[width=7cm, angle=0]{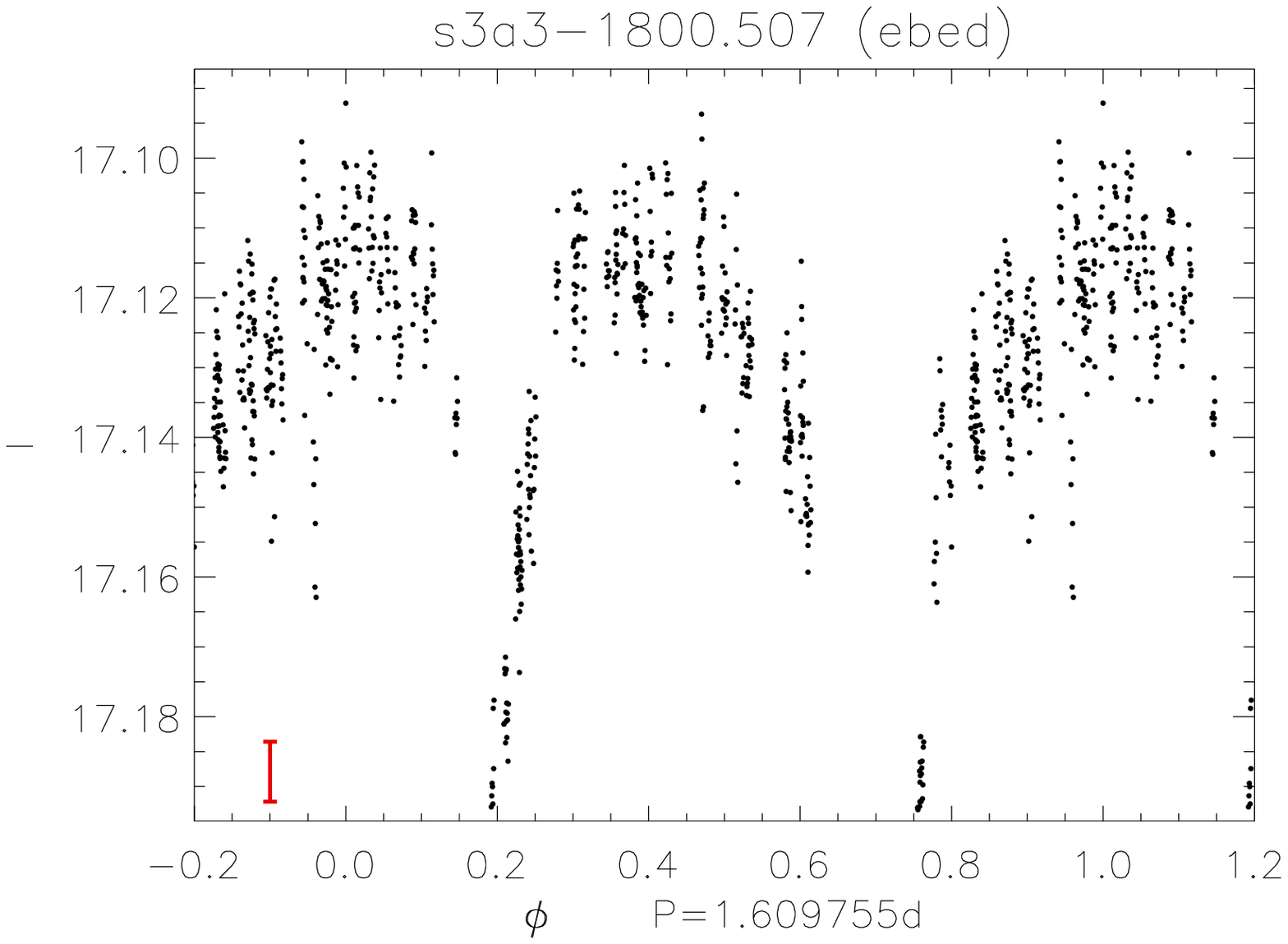}
\includegraphics[width=7cm, angle=0]{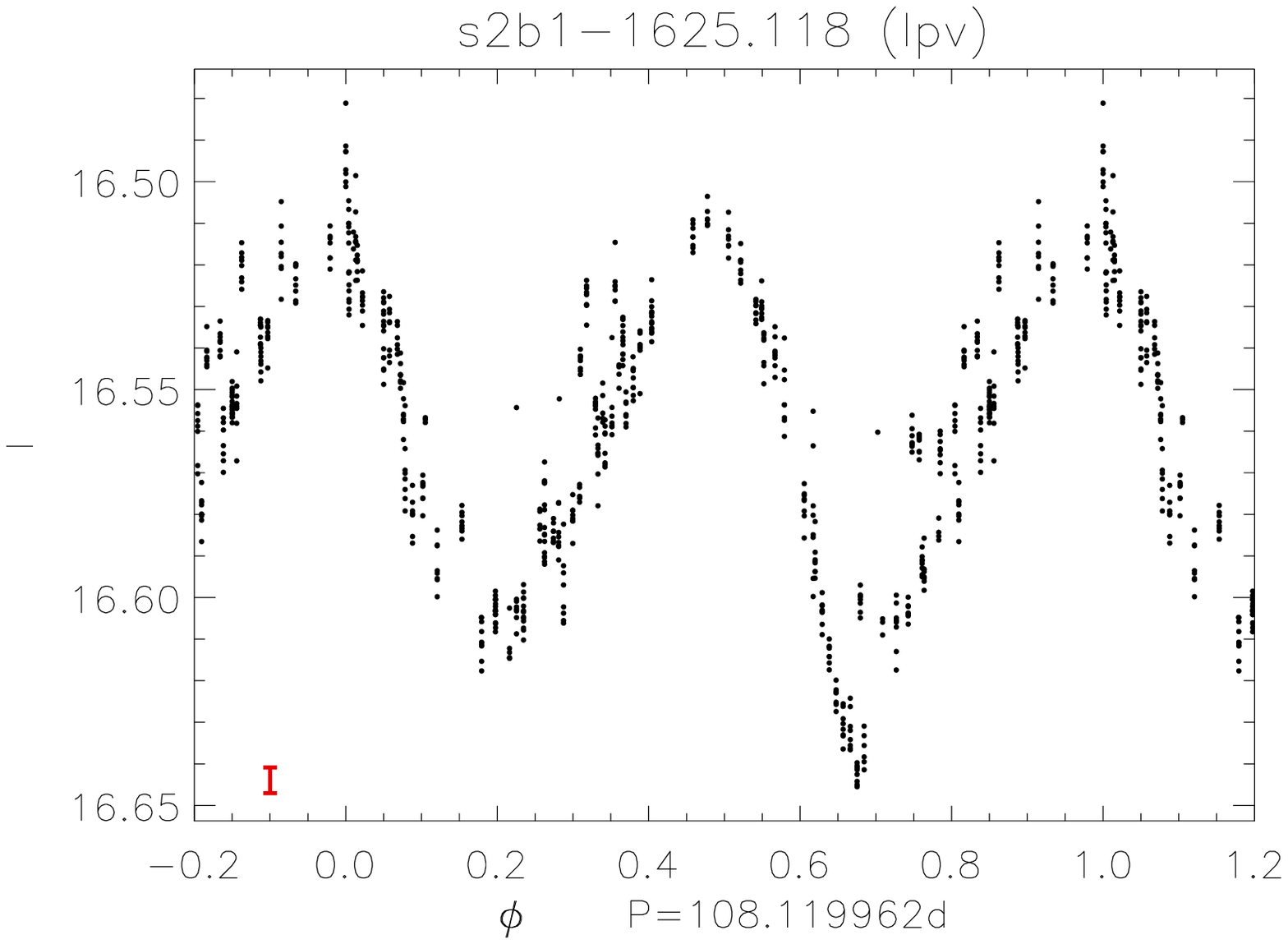}
\includegraphics[width=7cm, angle=0]{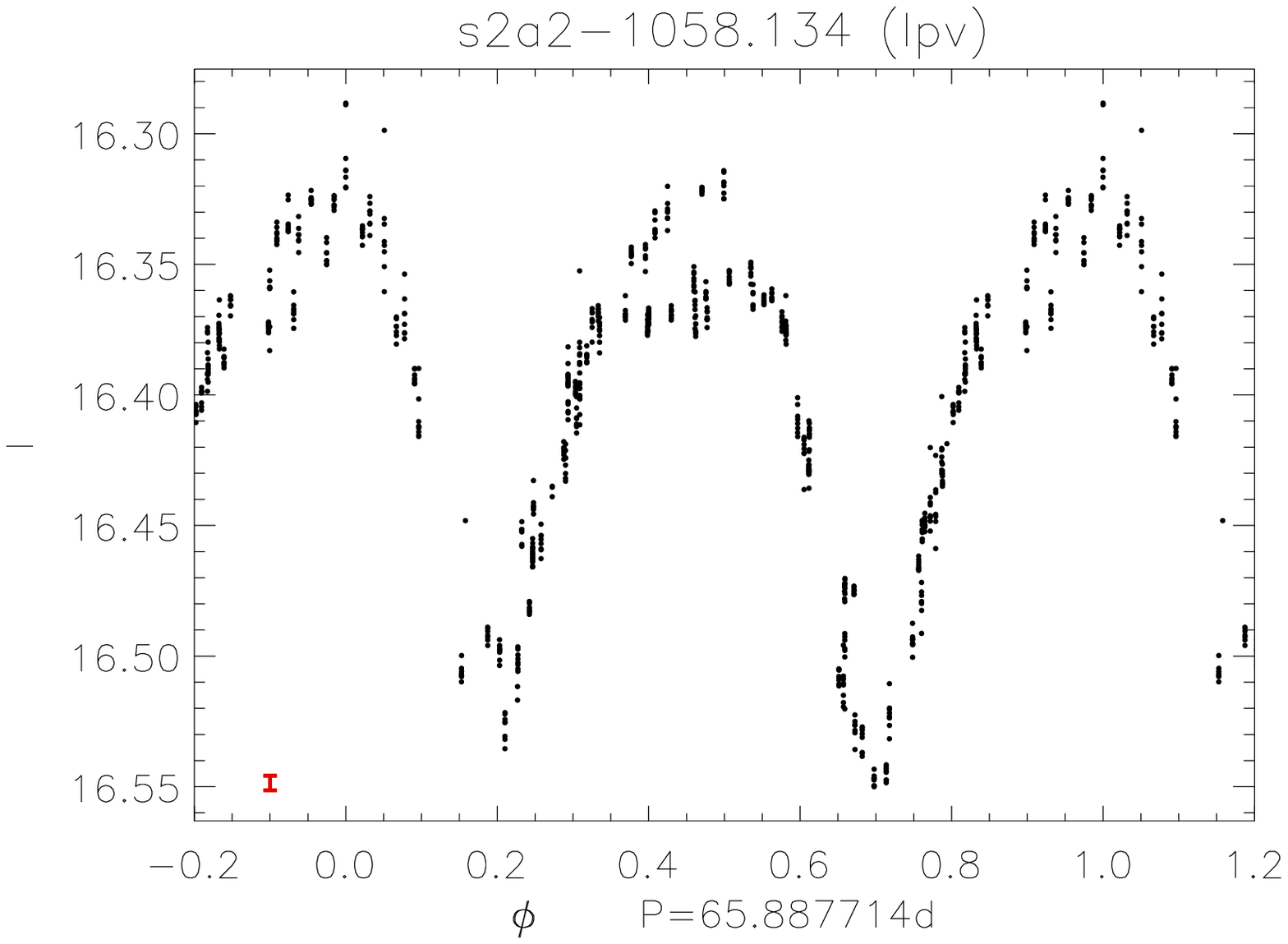}
\includegraphics[width=7cm, angle=0]{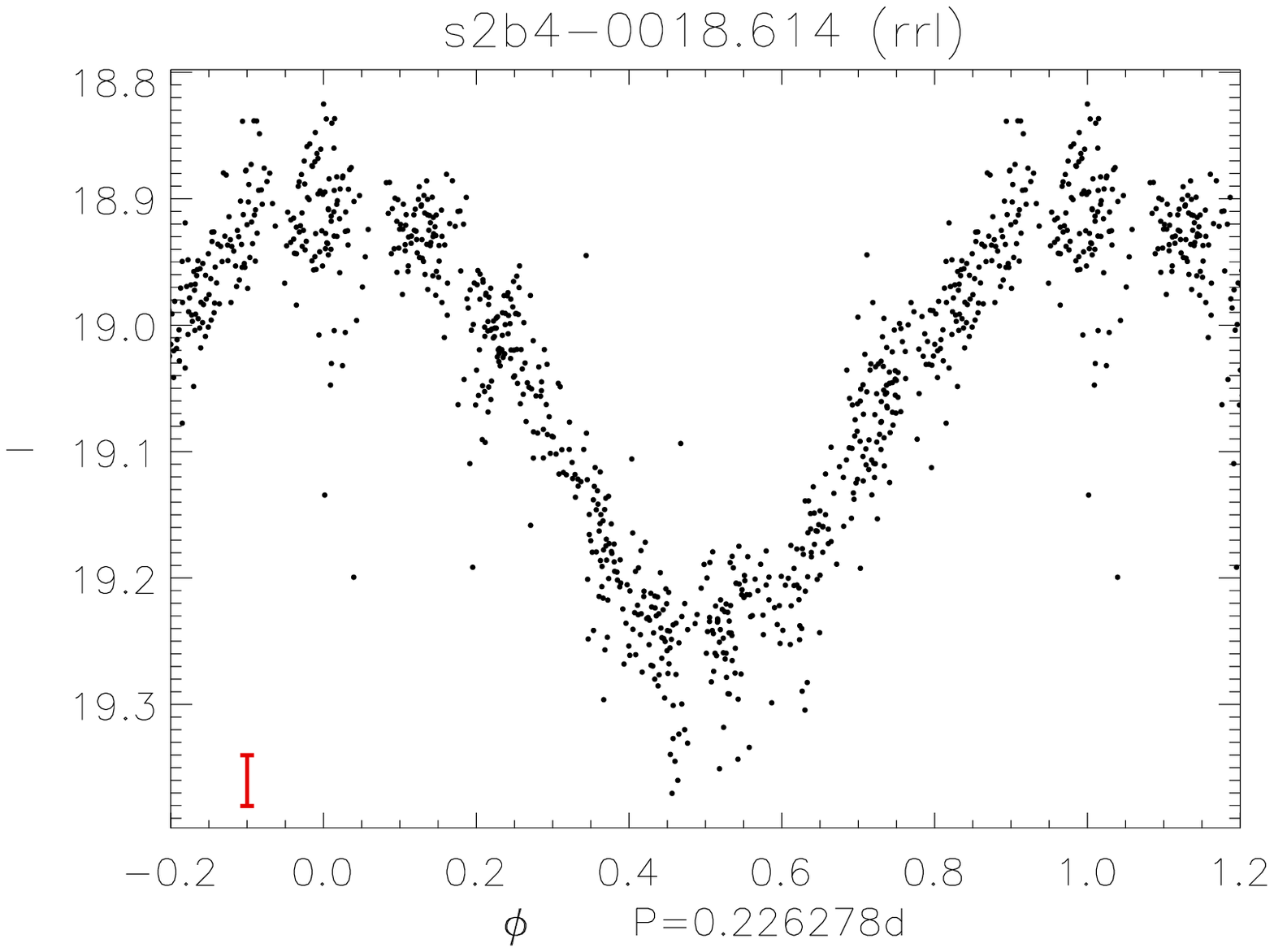}
\includegraphics[width=7cm, angle=0]{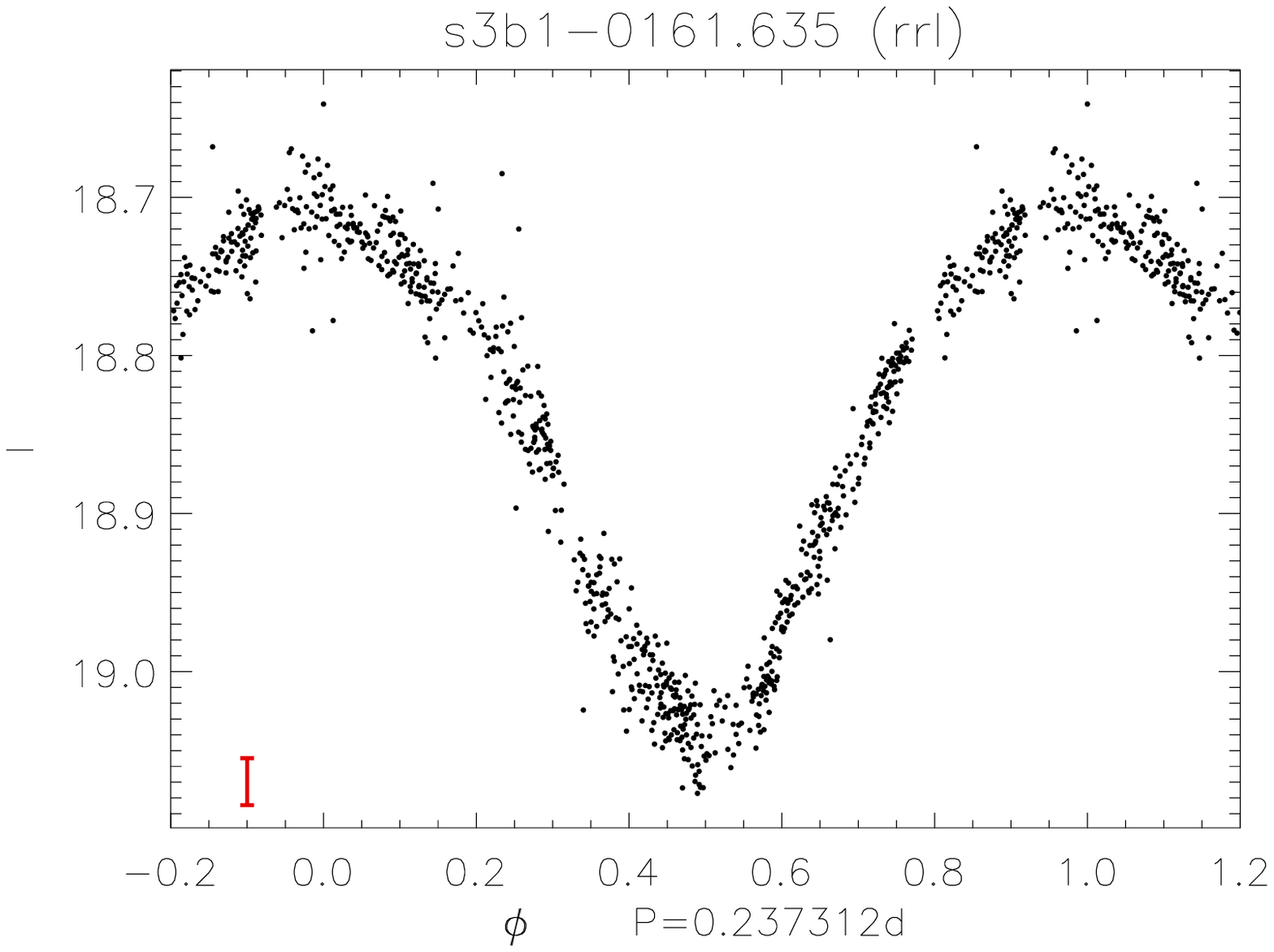}
\includegraphics[width=7cm, angle=0]{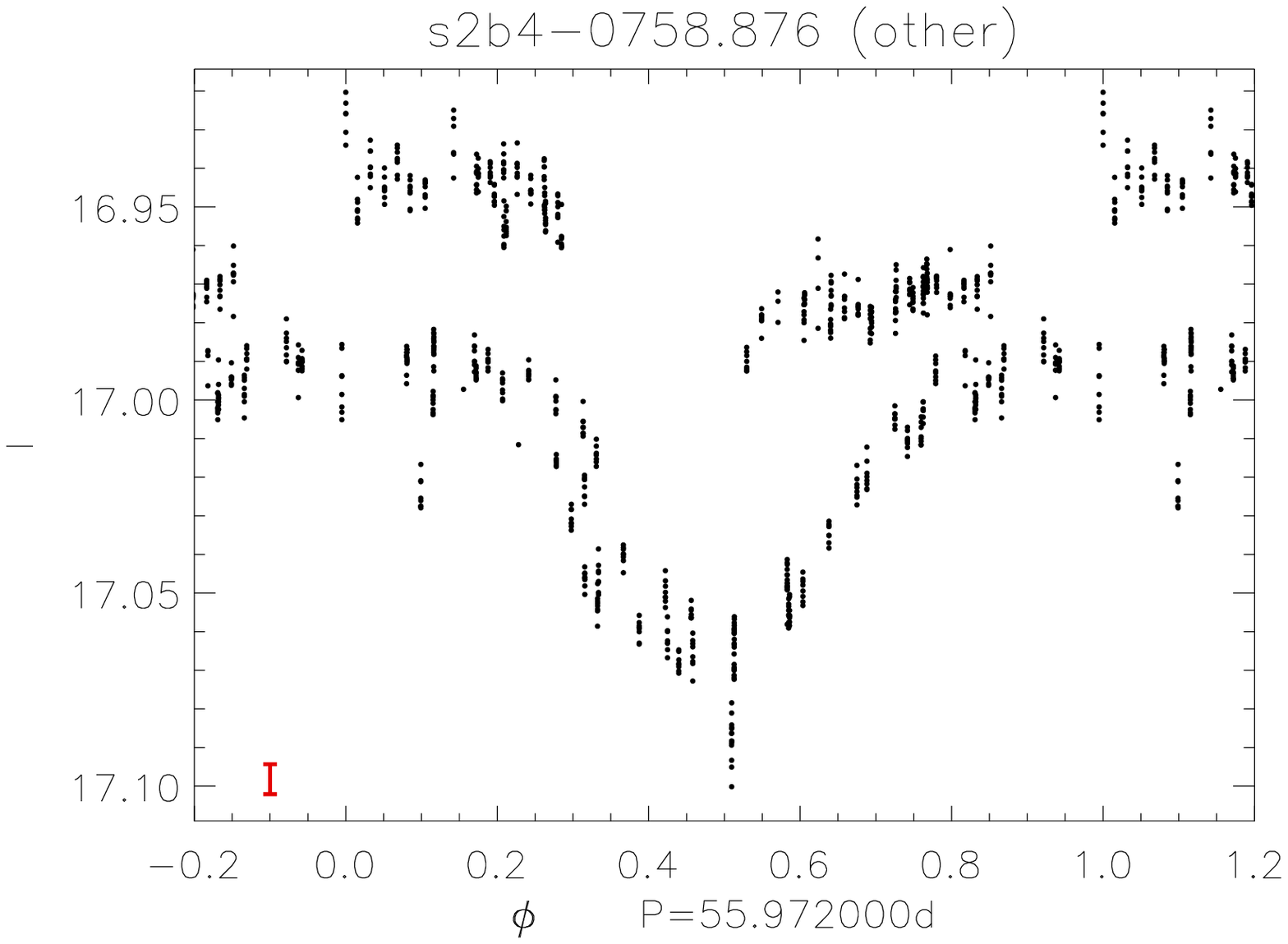}
\includegraphics[width=7cm, angle=0]{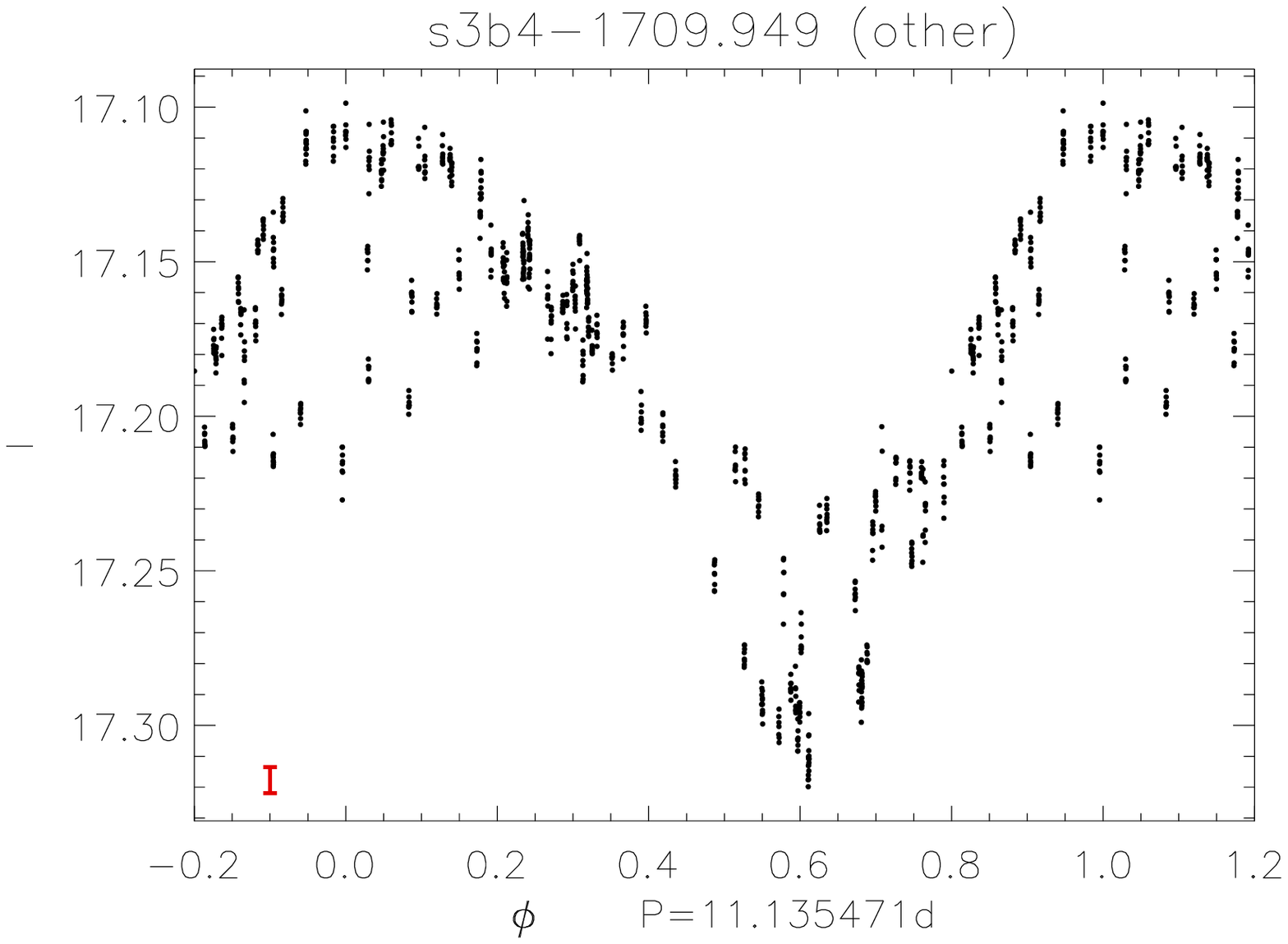}
\caption{Same as \Fig{fig:samplelc1}, for the detached eclipsing binaries, RR Lyrae, long-period variables, and variables we were unable to classify. \label{fig:samplelc2}}
\end{center}
\end{figure*}

Most of the new variables are either contact binaries (131), long-period variables (97 stars), or Cepheids (26 confirmed stars and 18 candidates), in addition to 6 detached eclipsing binaries, 2 Delta Scuti variables, 3 RR Lyrae, and 37 variables that did not fit any of the above classes. The new identified variables are shown on the $I,\ V - I$ CMD in \Fig{fig:cmdvar}. Sample light curves for each type of detected variables are shown in Fig. \ref{fig:samplelc1} and \ref{fig:samplelc2}. Their color and magnitude distribution covers the full observed color and magnitude ranges. In particular, the two new Delta Scuti stars (orange stars) are located at brighter magnitudes and bluer colors compared to the Galactic bulge turn-off location ($I \approx$ 18.7 and $V - I \approx$ 1.3 mag, see also \Fig{fig:cmd}), consistent with them being blue straggler stars as expected for this type of variables. The new confirmed (red circles) and candidate (purple stars) classical Cepheids are found in the 16.5 $< I <$ 19 and 1.5 $< V - I <$ 2.5 magnitude and color ranges in the left panel of \Fig{fig:cmdvar}. Only one Cepheid is much brighter in the group, having $I \approx$ 15 mag. 

The new RR Lyrae are marked with blue squares on the ($I,\ V - I$) CMD (one of the them does not have a $V$-band measurement). We assumed for the three RR Lyrae a metallicity of [Fe/H]$\approx$ -1.25 dex as the average metallicity of Galactic Bulge RR Lyrae \citep{kunder08}. To derive distances to our RR Lyrae, we then used the theoretical $I$-band period-luminosity-metallicity (PLZ) relation for first overtone RR Lyrae,

\begin{equation}
M_{I} = a + b\cdot \log(P)+c\cdot \mathrm{[Fe/H]}\,,
\end{equation}
\noindent
where $M_I$ is the absolute magnitude of the RR Lyrae, and the coefficients a, b and c are listed in Table~6 of \citet{marconi15}. We find distances of $\sim$ 27, 22, and 16 Kpc, respectively, with an uncertainty of $\sim$3 kpc. This includes the photometric calibration error, the error in the reddening estimate, and the error in the metallicity assumption. 

A more detailed analysis of the new classical Cepheids and RR Lyrae and their distance distribution will be presented in a forthcoming paper. 

The right panel of \Fig{fig:cmdvar} shows the new LPVs (green diamonds) and contact (black plus signs) and detached binaries (black triangles) on the $I,\ V - I$ CMD. Most of the identified binaries are located along the disk MS (blue plume) or the bulge MS. All the LPVs are distributed along the RGB phase, as we expect as they are either RGB or asymptotic giant branch stars. A study of the fainter counterpart of the population of binaries in the bulge was performed by using HST data and will be presented in the second paper of this series.

\begin{figure*}
\begin{center}
\includegraphics[width=10cm,angle=90]{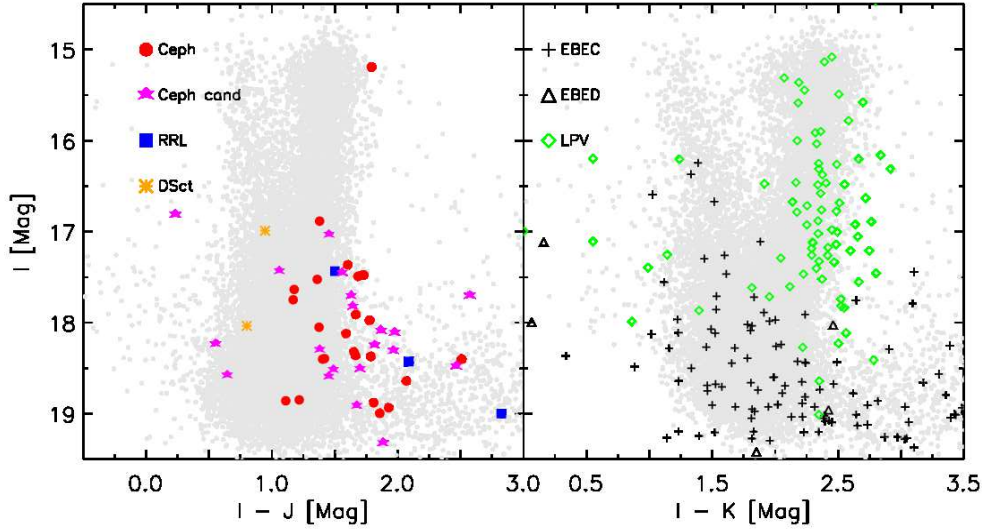}  \caption{Left: $I, I - J$ color-magnitude-diagram of stars observed in VIMOS field S1.A2 with the new identified RR Lyrae, Cepheids and Delta Scuti variables overplotted. Right: $I, I - K$ CMD with the new identified binaries and long period variables. Different variable classes are marked with different colors and labeled in the figures.\label{fig:cmdvarVVV}}
\end{center}
\end{figure*}

The identified variables have a similar distribution in the $I,\ I - J$ (left panel) and $I,\ I - K$ (right) CMDs of \Fig{fig:cmdvarVVV}. The color distribution of the variables is more spread in the CMDs that include near-IR filters due to a larger temperature sensitivity of the $I - J$ and $I - K$ colors compared to $V - I$.

%% =====================================================
%\begin{landscape}
\begin{table*}
\begin{center}
  \begin{tabular}{lccccccccccccccc}
\hline
ID	&Type	& Period	&  $<I>$   &$<V>$ &  $<J>$   &$<K>$   	\\
	&		&[d]	& [mag]	&[mag]	& [mag]	&[mag]	\\
\hline	   
s1a2-1012.243  &ceph  &2.920107  &18.37  &19.82  &-  &-  \\
s1a2-1051.225  &ceph  &4.321052  &18.40  &20.25  &16.87  &16.17  &\\
... \\
s1a2-1218.988  &cephcand  &14.497605  &18.58  &20.33  &16.17  &15.12  \\
s1a2-1708.053  &cephcand  &9.502042  &18.25  &-  &16.43  &15.63  \\
... \\
s1b1-1386.364  &dsct  &0.057131  &18.19  &19.26  &16.86  &16.24  \\
s2b1-0168.387  &dsct  &0.083012  &17.30  &18.56  &17.50  &16.29  \\
... \\
s1a2-0142.107  &ebec  &0.322285  &19.13  &-  &16.45  &15.44  \\
s1a2-0198.043  &ebec  &2.160771  &19.37  &-  &16.57  &15.76  \\
... \\
s1a2-1806.840  &ebed  &0.297271  &19.42  &-  &17.31  &16.90  \\
s1a3-0979.786  &ebed  &21.341234  &18.23  &20.21  &17.52  &17.02  \\
... \\
s1a2-0295.015  &lpv  &31.184000  &16.76  &18.69  &17.54  &16.26  \\
s1a2-0392.982  &lpv  &59.629198  &16.98  &18.98  &16.55  &15.52  \\
... \\
s1a2-0286.257  &other  &13.225344  &17.99  &20.02  &16.08  &15.32\\
s1a2-0913.144  &other  &15.372298  &18.30  &20.33  &17.08  &16.00\\
... \\
s2a2-1791.222  &rrl  &0.336787  &17.45  &19.00  &13.73  &12.79 \\
s2b4-0018.614  &rrl  &0.226278  &19.04  &20.90  &13.90  &12.79 \\
... \\
\hline
  \end{tabular}
  \caption{Except from the full table giving the main characteristics of the new variables. The variable types in divided into categories as described in the text: Cepheids (``ceph"), candidate Cepheids (``cephcand"), Delta Scuti (``dsct"), contact binaries (``ebec"), detached eclipsing binaries (``ebed"), long-period variables (``lpv"), RR Lyrae (``rrl"), and variables that did not fall into any of these categories (``other"). $J-$ and $K-$band magnitudes, when available, are taken from  VIRAC; for such stars, the corresponding ID from VIRAC is also given. The proper motion values from matches to both VIRAC and the \textit{Gaia} DR2 catalogue are also given, when available. For \textit{Gaia}, these are only given if a match was found within 2$\arcsec$ of a given star. Continued on Tab. \ref{tab:catalogue2}. \label{tab:catalogue1}}
  \end{center}
\end{table*}
%\end{landscape}

%% =====================================================

%% =====================================================
%\begin{landscape}
\begin{table*}
\begin{center}
  \begin{tabular}{lccccccccccccccc}
\hline
ID	&RA 	&Dec	&\textit{Gaia} ID&$\mu_{\alpha}$	&$\mu_{\delta}$	 &VVV ID &$\mu_{\alpha}$	&$\mu_{\delta}$\\
	&J2000.0	&J2000.0	&mas yr$^{-1}$ &mas yr$^{-1}$	\\
\hline	   
s1a2-1012.243  &17:59:15.75  &-29:10:46.7  &- &-  &- &165571... &-  &- \\
s1a2-1051.225  &17:59:01.04  &-29:10:54.6  &- &-0.57$\pm$0.50  &-5.83$\pm$0.40 &165568... &-  &- \\
... \\
s1a2-1218.988  &17:59:06.65  &-29:11:29.3  &- &-  &- &- &-  &- \\
s1a2-1708.053  &17:58:52.39  &-29:13:10.0  &- &-  &- &165690... &-  &- \\
... \\
s1b1-1386.364  &17:59:06.40  &-29:20:58.9  &406233... &-5.72$\pm$1.14  &-3.02$\pm$0.88 &- &-  &- \\
s2b1-0168.387  &17:59:45.31  &-29:15:07.1  &406235... &-  &- &165777... &-  &- \\
... \\
s1a2-0142.107  &17:59:12.70  &-29:07:47.2  &406234... &-1.49$\pm$0.36  &-4.31$\pm$0.30 &- &-  &- \\
s1a2-0198.043  &17:58:54.76  &-29:07:58.6  &406234... &-  &- &165770... &-  &- \\
... \\
s1a2-1806.840  &17:58:56.37  &-29:13:30.4  &- &-2.90$\pm$1.03  &-5.72$\pm$0.88 &- &-  &- \\
s1a3-0979.786  &17:58:07.97  &-29:10:36.7  &406234... &-  &- &- &-  &- \\
... \\
s1a2-0295.015  &17:59:07.43  &-29:08:18.7  &406234... &-  &- &- &-  &- \\
s1a2-0392.982  &17:58:58.48  &-29:08:38.8  &406234... &-3.71$\pm$0.75  &-6.97$\pm$0.61 &165770... &-  &- \\
... \\
s1a2-0286.257  &17:59:06.33  &-29:08:16.9  &- &-  &- &166039... &-  &- \\
s1a2-0913.144  &17:59:04.67  &-29:10:26.2  &406234... &-  &- &- &-  &- \\
... \\
s2a2-1791.222  &17:59:23.17  &-29:11:46.1  &406233... &-  &- &165941... &-3.31$\pm$1.77  &4.44$\pm$1.75 \\
s2b4-0018.614  &17:58:49.07  &-29:14:35.1  &406234... &-7.29$\pm$0.17  &-4.94$\pm$0.14 &- &-  &- \\
... \\
\hline
  \end{tabular}
  \caption{Continued from Tab. \ref{tab:catalogue1}. \label{tab:catalogue2}}
  \end{center}
\end{table*}
%\end{landscape}

%% =====================================================

\begin{figure}
\begin{center}
\includegraphics[width=8cm, angle=0]{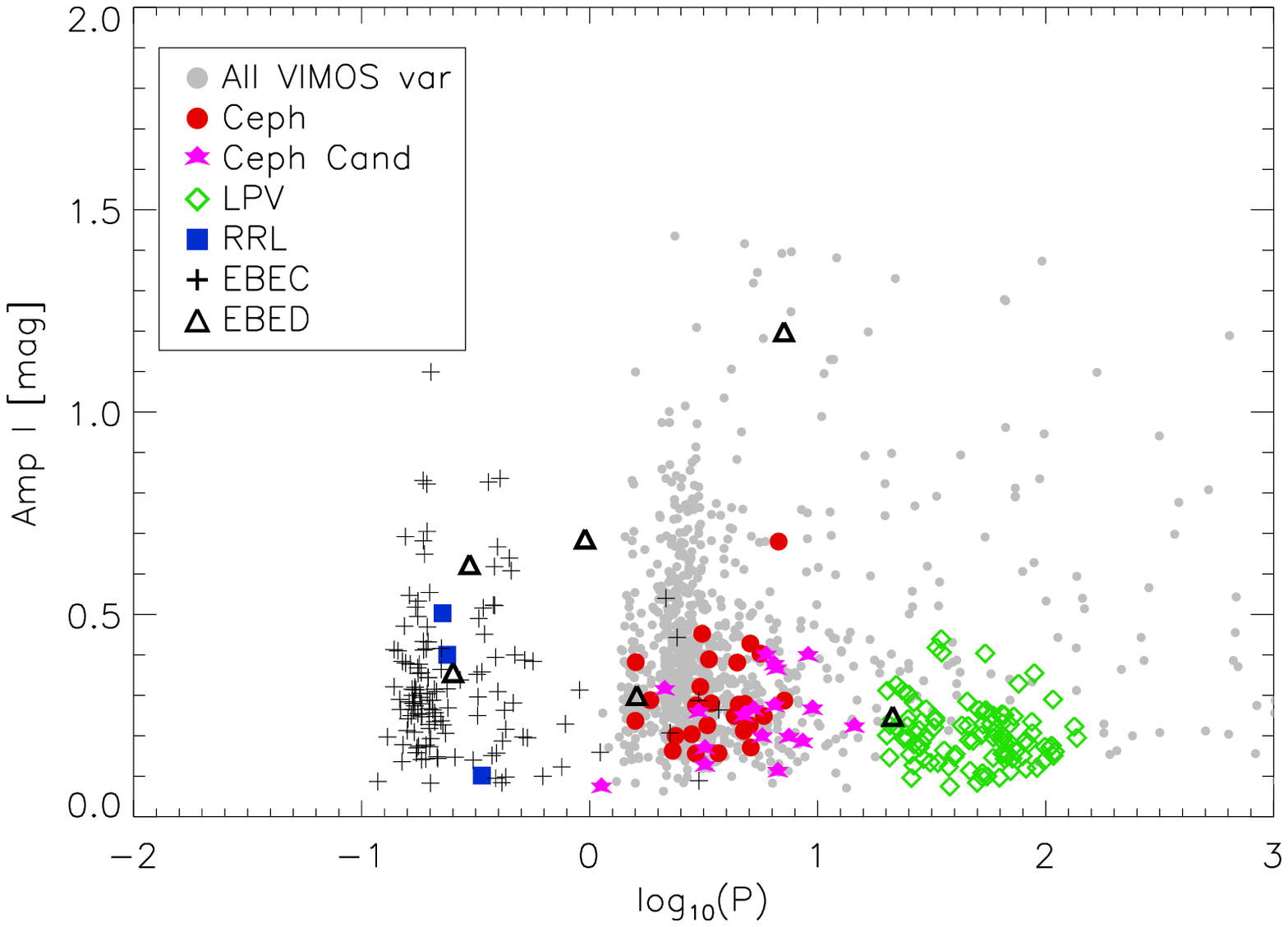}  
\caption{Bailey diagram, showing log($P$) vs. $I-$band amplitude for the newly detected variables in our sample. Different classes of variables are plotted with different colours and symbols, showing clear separation by variable type. \label{fig:bailey}}
\end{center}
\end{figure}

Together with PCA and Fourier parameters'  ratios shown in \ref{fig:lca}, the Bailey Diagram can also be used as a distance-independent, reddening-free 
diagnostic for differentiating among the variability classes.  As a diagram of the luminosity amplitude versus the pulsation period, it is indeed readily legible and works as a starting point for the other classifications criteria.
For instance, it is well known that Classical Cepheids present a typical ``double-peak'' distribution on this diagram, with
a linear increase in amplitude for periods shorter and longer than the so-called Hertzsprung progression, located at $\sim$10 days for Milky Way Cepheids \citep{bono02,inno15,bhardwaj17}.

In \Fig{fig:bailey}, we show the Bailey diagram for the different classes we defined. 
The linear-increasing regime for the selected Cepheids is clearly visible for periods between 4 and 10 days, 
while the second peak does not appear, as we did not identified candidates with longer periods. This is because they would be too bright ($I$-band mag $\lesssim$ 13) and therefore saturated in our survey. However, we find variables classified as LPV in our sample at longer periods. In fact, they show a rather low amplitudes with respect to the one expected for Cepheids at these periods.

\section{Conclusions}\label{sec:conclusions}
In this paper, we have presented a sample of 320 newly identified variables towards the Galactic Bulge, including many contact and eclipsing binaries, and a significant number of Cepheids. The catalogue of variables presented here is made available electronically, and will enable future studies making use of variability to probe Galactic structure. The Cepheids we have identified will be particularly useful for such purposes. Some of the long-period variables we presented may also be consistent with being Cepheid pulsators, but further observations, for instance in the form of spectroscopic follow-up, are required to confirm this. Should some of these stars be confirmed as Cepheids, period-luminosity relations would place them over 200 kpc from the Galactic Center, further than populations of currently known Cepheids, opening up interesting insights about the structure of the Milky Way.

\section*{Acknowledgements}

The authors thank Leigh Smith and Phil Lucas for help with the use of VIRAC proper motion catalogs. This research has made use of the SIMBAD database, operated at CDS, Strasbourg, France. NK acknowledges funding from HST grants GO-120586, GO-13057, and GO-13464 (PI: Sahu). MZ acknowledges support by the Ministry of Economy, Development, and Tourism's Millennium Science Initiative through grant IC120009, awarded to The Millennium Institute of Astrophysics (MAS), by Fondecyt Regular 1150345 and by the BASAL-CATA Center for Astrophysics and Associated Technologies PFB-06. LI acknowledges support by the Sonderforschungsbereich SFB 881
"The Milky Way System" (subproject A3) of the German Research Foundation (DFG). FSM acknowledges the support by the DFG Cluster of Excellence "Origin and Structure of the Universe". This work has made use of data from the European Space Agency (ESA) mission
{\it Gaia} (\url{https://www.cosmos.esa.int/gaia}), processed by the {\it Gaia}
Data Processing and Analysis Consortium (DPAC,
\url{https://www.cosmos.esa.int/web/gaia/dpac/consortium}). Funding for the DPAC
has been provided by national institutions, in particular the institutions
participating in the {\it Gaia} Multilateral Agreement.

%%%%%%%%%%%%%%%%%%%%%%%%%%%%%%%%%%%%%%%%%%%%%%%%%%



\bibliographystyle{mnras}
\bibliography{thesisbib,calamida} % if your bibtex file is called example.bib
% \bibliography{calamida}
% * <noekains@gmail.com> 2016-11-15T17:34:29.313Z:
%
% ^.

%%%%%%%%%%%%%%%%%%%%%%%%%%%%%%%%%%%%%%%%%%%%%%%%%%

%%%%%%%%%%%%%%%%% APPENDICES %%%%%%%%%%%%%%%%%%%%%

%\appendix


%%%%%%%%%%%%%%%%%%%%%%%%%%%%%%%%%%%%%%%%%%%%%%%%%%

% Don't change these lines
\bsp	% typesetting comment
\label{lastpage}
\end{document}